\documentclass[twocolumn, trackchanges]{aastex631}

\usepackage{amsmath}

\newcommand{\Ab}{\boldsymbol{A}}
\newcommand{\Bb}{\boldsymbol{B}}
\newcommand{\Ap}{\boldsymbol{A}_{\rm p}}
\newcommand{\Bp}{\boldsymbol{B}_{\rm p}}
\newcommand{\Aj}{\boldsymbol{A}_{\rm J}}
\newcommand{\Bj}{\boldsymbol{B}_{\rm J}}
\newcommand{\ebt}{\boldsymbol{e}_t}
\newcommand{\ebr}{\boldsymbol{e}_r}
\newcommand{\ebph}{\boldsymbol{e}_\phi}
\newcommand{\ebth}{\boldsymbol{e}_\theta}
\newcommand{\nb}{\boldsymbol{n}}
\newcommand{\xb}{\boldsymbol{x}}

\newcommand{\vb}{\boldsymbol{v}}
\newcommand{\bfzero}{\boldsymbol{0}}
\newcommand{\bnabla}{\boldsymbol{\nabla}}
\newcommand{\ddy}[2]{\frac{\partial #1}{\partial #2}}
\newcommand{\dl}{d\boldsymbol{l}}

\renewcommand{\edit}[1]{#1}

\accepted{December 23, 2025}

\submitjournal{ApJ}

\begin{document}

\title{Energy Bounds from Relative Magnetic Helicity in Spherical Shells}

\author[0000-0002-2728-4053]{Anthony R. Yeates}
\affiliation{Department of Mathematical Sciences, Durham University, Durham, DH1 3LE, UK}
\author[0000-0003-0981-2761]{Gunnar Hornig}
\affiliation{Division of Mathematics, University of Dundee, Dundee, DD1 4HN, UK}

\correspondingauthor{Anthony R. Yeates}
\email{anthony.yeates@durham.ac.uk}

\begin{abstract}
Relative magnetic helicity is commonly used in solar physics to avoid the well known gauge ambiguity of standard magnetic helicity in magnetically open domains. But its physical interpretation is difficult owing to the invocation of a reference field. For the specific case of spherical shell domains (with potential reference field), relative helicity may be written intrinsically in terms of the magnetic field alone, without the need to calculate the reference field or its vector potential. We use this intrinsic expression to prove that non-zero relative helicity implies lower bounds for both magnetic energy and free magnetic energy, generalizing the important Arnol'd inequality known for closed-field magnetic helicity. Further, we derive a stronger energy bound by spatially decomposing the relative helicity over a magnetic partition of the domain to obtain a new ideal invariant which we call unsigned helicity. The bounds are illustrated with analytical linear force-free fields (that maximize relative helicity for given boundary conditions) as well as a non-potential data-driven model of the solar corona. These bounds confirm that both relative helicity and the unsigned helicity can influence the dynamics in the solar corona.
\end{abstract}


\section{Introduction} \label{sec:intro}

In solar physics, it is common to avoid gauge ambiguities in the classical magnetic helicity,
\begin{equation}
H = \int_V\Ab\cdot\Bb\,dV,
\label{eq:h}
\end{equation}
by measuring instead the relative (magnetic) helicity of \citet{berger1984relative-helicity}, which we denote $H_{\rm R}$. This is most commonly computed from the \citet{finn1985helicity} formula,
\begin{equation}
    H_{\rm R} = \int_V(\Ab + \Ap)\cdot(\Bb - \Bp)\,dV,
    \label{eq:hrfinn}
\end{equation}
where $\Bb=\bnabla\times\Ab$ is the magnetic field of interest, and $\Bp=\bnabla\times\Ap$ is a reference field matching $\nb\cdot\Bp|_{\partial V}=\nb\cdot\Bb|_{\partial V}$ on the boundary $\partial V$. Using \eqref{eq:hrfinn}, the relative helicity is easily shown to be invariant under gauge transformations of either $\Ab$ or $\Ap$. It is conserved in ideal magnetohydrodynamics in line-tied situations where $\vb=\bfzero$ on $\partial V$, and -- crucially -- remains approximately conserved even under magnetic reconnection \citep{berger1984dissipation, blackman2015ssr}. We use the notation $\Bp$  because the reference field is almost universally taken to be the unique potential (current-free) field satisfying $\bnabla\times\Bp=\bfzero$ in $V$ as well as the boundary matching condition. Among all magnetic fields satisfying these boundary conditions, $\Bp$ is well known to have minimum energy, or in other words to minimize the $L^2$-norm $\|\Bb\|^2 \equiv \int_V|\Bb|^2\,dV$. By definition, $H_{\rm R}$ vanishes if $\Bb=\Bp$. The relative helicity $H_{\rm R}$ has been widely used in solar physics to analyze magnetic activity \citep[e.g.,][]{demoulin2009review, toriumi2019lrsp, thalmann2021issi4}.

Oddly, much less attention has been paid in the literature to what $H_{\rm R}$ is actually measuring, and why it can reasonably be expected to give us useful information about the dynamics. In particular, the fact that $H_{\rm R}$ is conserved (or approximately conserved) in a line-tied evolution is not in itself a particularly compelling reason to study it, since any choice of classical helicity $H$ is also conserved in such an evolution, provided $\nb\times\Ab|_{\partial V}$ is held fixed over time. Indeed, as long as $\Bb$ contains at least one open magnetic field line (intersecting $\partial V$), one can choose the gauge of $\Ab$ to give $H$ an arbitrary numerical value, of either sign. This value will remain unchanged in the evolution provided that line-tying holds and that the gauge of $\nb\times\Ab|_{\partial V}$ is fixed. Thus although this $H$ is conserved, its numerical value cannot tell us anything about the magnetic field, since it is arbitrary.

By contrast, in the magnetically-closed case, both the sign and magnitude of $H$ are meaningful: it is gauge independent and measures the average pairwise linking between infinitesimal flux tubes within the magnetic field \citep[e.g.,][]{moffatt1992,berger1999, moffatt2019book}. Moreover, for this case, $H$ is well known to obey an inequality of the form
\begin{equation}
    |H| \leq C\|\Bb\|^2,
    \label{eq:arnold}
\end{equation}
showing that a non-zero helicity puts a lower bound on the magnetic energy \citep{arnold1986hopf, cantarella2001bounds, berger2003topologicalmhd}. Our main aim in this paper is to show that an analogous inequality holds for $H_{\rm R}$, at least when $V$ is a spherical shell (the region between two concentric spheres) and the reference field is potential.

In realistic magnetic fields, the bound $|H|/C$ from \eqref{eq:arnold} can be a significant underestimate for $\|\Bb\|^2$. For example, it is certainly possible to have $H=0$ even though the magnetic field has non-trivial topology \citep[e.g.,][]{candelaresi2011knots, pontin2016braids}. Stronger energy bounds have been derived in terms of so-called crossing numbers, as reviewed by \citet{berger2003topologicalmhd}. These apply to both magnetically-closed knotted fields \citep{freedman1991} and ``braided'' fields between two boundaries \citep{berger1993braids, aly2014braids, yeates2014bianchi}. However, these crossing numbers are not conserved in an ideal evolution. In this paper, we focus on obtaining energy inequalities where the left-hand side is also a conserved quantity. This is because our overall aim is not to generate the tightest possible bound on energy for a known magnetic field (indeed, if $\Bb$ is known then one can compute the energy directly), but to (i) put robust bounds on how energy could evolve, and (ii) show that $H_{\rm R}$ (and related conserved quantities) can constrain the dynamics. With the crossing number approach, it is a challenging problem to determine the minimum possible energy over all ideally accessible configurations \citep[but see, e.g.][]{ricca2013}. Instead, we will show (Section \ref{sec:decomp}) how to strengthen the inequality while retaining ideal invariance of the left-hand side, by partitioning $V$ into magnetic subdomains.

The reason for restricting to a spherical shell is that we can take advantage of poloidal-toroidal decomposition of $\Bb$ \citep{backus1958-PT, berger1985lfff}. As will be shown in Section \ref{sec:sph}, the known Green's function expressions for the poloidal and toroidal potentials give us an explicit expression for $H_{\rm R}$ in terms of $\Bb$ alone, with no reference to either $\Bp$ or to any vector potentials $\Ab$ or $\Ap$. Analogous expressions are available in the planar slab (region between two parallel planes), but in other domains a generalized decomposition is required \citep{berger2018poloidal, yi2022poloidal}; it is quite possible that $H_{\rm R}$ may still give a lower bound for $\|\Bb\|^2$, but our proofs do not apply directly.

In spherical shells, the simplification of $H_{\rm R}$ has previously been exploited to relate energy to helicity for the specific case of force-free fields \citep[up to a boundary term;][]{berger1988flh}, to interpret $H_{\rm R}$ as a sum of pairwise linking between elementary (open-ended) flux tubes \citep{demoulin2006mutualhelicity}, and to relate helicity to pairwise winding numbers between field lines \citep{xiao2023spherical}. The latter two works provide physical interpretations of $H_{\rm R}$ without recourse either to the reference field $\Bp$, or to fields external to $V$ \citep[unlike][]{berger1984relative-helicity, schuck2024linkages}. This paper will provide further ``intrinsic'' evidence for the dynamical significance of $H_{\rm R}$ in spherical shells.

The paper is organized as follows. Section \ref{sec:sph} reviews the fundamental expression for $H_{\rm R}$ on a spherical shell, on which the rest of the paper is based. The main inequality giving $|H_{\rm R}|\leq C\|\Bb\|^2$ is presented in Section \ref{sec:energy}, where we also obtain an analogous bound replacing $\|\Bb\|^2$ by free magnetic energy $\|\Bb-\Bp\|^2$. Section \ref{sec:decomp} then derives the improved, partition-based bound for $\|\Bb\|^2$ where $H_{\rm R}$ is replaced by so-called ``unsigned helicity''. The utility of this improved bound is illustrated in Section \ref{sec:eg} using a numerical model of the solar corona, with conclusions in Section \ref{sec:conclusion}.

\section{Spherical Shells} \label{sec:sph}

In this paper, $V$ denotes a spherical shell with radius $r\in[r_0, r_1]$. The spherical surfaces of constant $r$ are labelled $S_r$, with $S_0\equiv S_{r_0}$ denoting the inner boundary and $S_1\equiv S_{r_1}$ the outer boundary. We assume no net magnetic flux or current through $S_r$ for any $r\in[r_0, r_1]$, meaning
\begin{equation}
    \oint_{S_r}\ebr\cdot\Bb\,dS = \oint_{S_r}\ebr\cdot\bnabla\times\Bb\,dS = 0.
    \label{eq:nonet}
\end{equation}

\subsection{Poloidal-Toroidal Vector Potential}

In our spherical shell with \eqref{eq:nonet}, we can write $\Bb=\bnabla\times\Ab^*$ where
\begin{equation}
    \Ab^* = \bnabla\times(P\ebr) + T\ebr,
    \label{eq:ptA}
\end{equation}
 which is the well-known poloidal-toroidal decomposition of $\Bb$ \citep{backus1958-PT}, discussed in the helicity context by \citet{berger1985lfff} and \citet{berger2018poloidal}.

To see that $\Ab^*$ exists, note that taking the curl of \eqref{eq:ptA} once or twice gives
\begin{align}
    \nabla^2_hP &= -\ebr\cdot\Bb,\label{eq:lapP}\\
    \nabla^2_hT &= -\ebr\cdot\bnabla\times\Bb,\label{eq:lapT}
\end{align}
on each $S_r$, in terms of the surface Laplacian
\begin{equation}
    \nabla^2_h \equiv \frac{1}{r^2\sin\theta}\left[\ddy{}{\theta}\left(\sin\theta\ddy{}{\theta}\right) + \frac{1}{\sin\theta}\ddy{^2}{\phi^2}\right].
\end{equation}
These two Poisson equations are solvable for $P$ and $T$ thanks to our assumptions \eqref{eq:nonet}. Note that the solutions are unique only up to additive functions of $r$. Neither of these ambiguities affect $H_{\rm R}$, however changing $T$ does change $\Ab^*$. To derive our results in Section \ref{sec:intrinsicA} and beyond, we will need to fix $\Ab^*$ by choosing the particular $T$ with
\begin{equation}
    \oint_{S_r}T\,dS = 0 \quad \textrm{for each } r\in [r_0, r_1].
    \label{eq:tgauge}
\end{equation}

\subsection{Simple Expression for Relative Helicity}

It is known that the expression \eqref{eq:hrfinn} for $H_{\rm R}$ simplifies considerably in our spherical shell geometry \citep{berger1988flh, devore2000diffrot, demoulin2007review}. To see this, note that applying the equivalent decomposition \eqref{eq:ptA} to the reference vector potential gives $\Ap^*=\bnabla\times(P_{\rm p}\ebr\big)$, because $\bnabla\times\Bp=\bfzero$. Then integrating by parts gives
\begin{align}
    \int_V\Ap^*\cdot\Bp\,dV &= \int_V\Bp\cdot\bnabla\times\big(P_{\rm p}\ebr\big)\,dV = 0.
\end{align}
From \eqref{eq:hrfinn}, this leaves
\begin{align}
    H_{\rm R} &= \int_V\Ab^*\cdot\Bb\,dV + \int_V(\Ap^*\cdot\Bb - \Ab^*\cdot\Bp)\,dV\\
    &= \int_V\Ab^*\cdot\Bb\,dV + \oint_{\partial V}\nb\cdot(\Ab^*\times\Ap^*)\,dS.
\end{align}
But since $P$ and $P_{\rm p}$ both solve the same Poisson equation \eqref{eq:lapP}, they can differ only by an additive constant on each of $S_0$, $S_1$, so that the boundary integral vanishes. Therefore
\begin{equation}
    H_{\rm R} = \int_V\Ab^*\cdot\Bb\,dV.
    \label{EQ:HPT}
\end{equation}
In other words, for the spherical shell, $H_{\rm R}$ is equivalent to $H$ in the poloidal-toroidal gauge. Note that this expression holds for any choice of $T$. It is the starting point for the energy bounds derived in this paper, for which we will use the particular choice \eqref{eq:tgauge} for $T$.

\subsection{Intrinsic Expression for $\Ap^*$} \label{sec:intrinsicA}

\begin{figure}
    \centering
    \includegraphics[width=\linewidth]{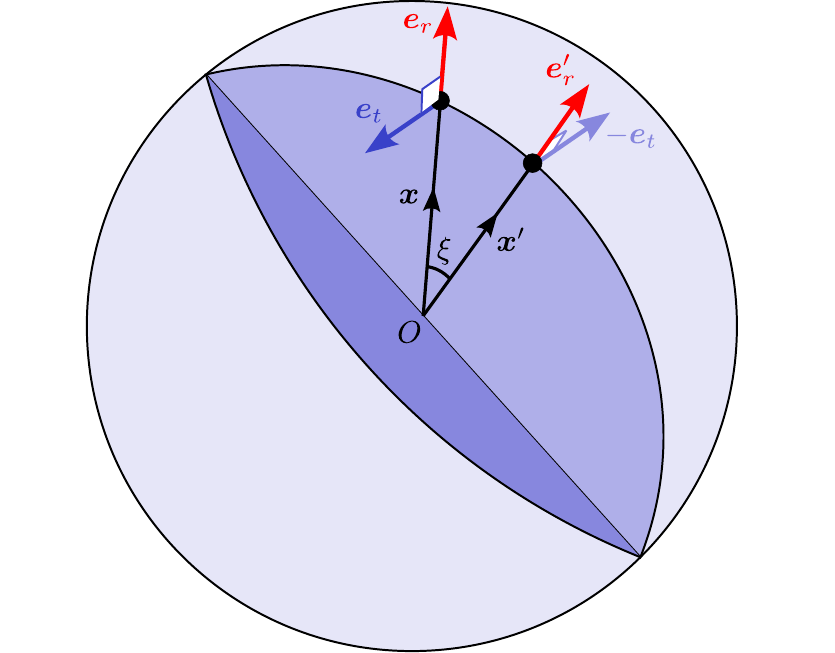}
    \caption{Geometry for Sections \ref{sec:intrinsicA} and \ref{sec:intrinsicHR} -- in particular, the intrinsic expression \eqref{eq:hwinding} for $H_R$. Note that $\xb=r\ebr$ and $\xb'=r\ebr'$.}
    \label{fig:sph-winding}
\end{figure}

To derive our energy bounds, we will express $\Ab^*$ in terms of $\Bb$. This can be done using Green's functions to write the solutions to \eqref{eq:lapP} and \eqref{eq:lapT} as
\begin{align}
    P(\xb) &= -\int_{S_r'}(\ebr'\cdot\Bb)(\xb')G(\xb,\xb')\,dS',\label{eq:greenP}\\
    T(\xb) &= -\int_{S_r'}(\ebr'\cdot\bnabla'\times\Bb)(\xb')G(\xb,\xb')\,dS'. \label{eq:greenT}
\end{align}
Here $\xb=r\ebr$ and $\xb'=r\ebr'$ are points on the sphere $S_r$ and the spherical Green's function is
\begin{equation}
    G(\xb,\xb') = \frac{1}{4\pi}\log(1-\cos\xi),
\end{equation}
with $\xi$ the angle between $\xb$ and $\xb'$ (see Figure \ref{fig:sph-winding}). The particular solutions \eqref{eq:greenP} and \eqref{eq:greenT} both have zero mean on every $S_r$ \citep[e.g.][]{backus1958-PT}, so in particular \eqref{eq:greenT} satisfies  \eqref{eq:tgauge}. Substituting into \eqref{eq:ptA} gives
\begin{align}
    \Ab^*(\xb) &= \ebr\times\left(\int_{S_r'}B_r(\xb')\bnabla_h G\,dS'\right) \nonumber\\ &+ \left(\int_{S_r'}\Bb(\xb')\cdot\ebr'\times\bnabla_h'G\,dS'\right)\ebr.
    \label{eq:agreens}
\end{align}
As observed by \citet{xiao2023spherical}, we have
\begin{equation}
    \ebr\times\bnabla_h G = \frac{\xb'\times\xb}{4\pi r^3(1-\cos\xi)},
    \label{eq:gradg}
\end{equation}
and $\ebr'\times\bnabla_h'G = -\ebr\times\bnabla_h G$.

\subsection{Intrinsic Expression for $H_{\rm R}$} \label{sec:intrinsicHR}

As an aside, we can use \eqref{eq:agreens} to derive a new expression for $H_{\rm R}$. Following \citet{xiao2023spherical}, we define the unit vector
\begin{equation}
    \ebt= \frac{\xb'\times\xb}{|\xb'\times\xb|} = \frac{\xb'\times\xb}{r^2\sin\xi},
\end{equation}
which is orthogonal to the great circle connecting $\xb$ and $\xb'$ (Figure \ref{fig:sph-winding}).  Substituting \eqref{eq:agreens} into \eqref{EQ:HPT} and defining $\Gamma(\xi)=\sin\xi/(1-\cos\xi)$ leads to the alternative expression
\begin{align}
    H_{\rm R} &= \int_{r_0}^{r_1}\int_{S_r\times S_r'}\frac{\Gamma(\xi)}{4\pi r}\Big[B_r(\xb')B_t(\xb) \nonumber\\
    &\qquad- B_r(\xb)B_t(\xb')\Big]\,dS'\,dS\,dr,
    \label{eq:hwinding}
\end{align}
where $B_t \equiv \Bb\cdot\ebt$. Note that $\Gamma(\xi)$ has a singularity when $\xb'=\xb$, since \edit{$\sin\xi/(1-\cos\xi)\sim 2/\xi$} as $\xi\to 0$. However, choosing coordinates for $S'$ where $\xb$ is the north pole makes $\xi$ into colatitude, so the surface element $dS'$ contributes another $\sin\xi$, showing that the integrand remains finite. Expression \eqref{eq:hwinding} is ``intrinsic'' in the sense that it refers to neither any vector potential nor any reference field.

From Figure \ref{fig:sph-winding}, we see that $B_t(\xb)$ is the component of $\Bb(\xb)$ ``rotating around'' $\xb'$, and \textit{vice versa} for $B_t(\xb')$. So \eqref{eq:hwinding} has the clear interpretation of measuring the average winding of $\Bb$ between all pairs of points, with respect to the radial direction. \citet{xiao2023spherical} show that the weighting $\Gamma(\xi)$ accounts naturally for the spherical geometry; this factor vanishes when $\xi=\pi$, meaning $\xb$ and $\xb'$ are antipodal. By recasting \eqref{eq:hwinding} as an integral over field lines rather than height, they also show how to express $H_{\rm R}$ in a spherical shell as a flux-weighted average of pairwise winding numbers between magnetic field lines. This is analogous to the topological interpretation of $H$ in terms of linking numbers.

We could also view the expression \eqref{eq:hwinding} for $H_{\rm R}$ as a two-point correlation function for $\Bb$, using off-diagonal components of the tensor $B_iB_j$. This is reminiscent of the well-known interpretation for $H$ in periodic domains (using Coulomb gauge, $\nabla\cdot\Ab^{\rm c}=0$), used to evaluate helicity in studies of solar wind turbulence \citep{matthaeus1982solarwind,narita2024review}.

\section{Global Energy Bounds} \label{sec:energy}

In Section \ref{sec:energy1} we generalize the \citet{arnold1986hopf} inequality \eqref{eq:arnold} to $H_{\rm R}$, for a spherical shell. Before doing so, recall the typical proof for \eqref{eq:arnold} in a magnetically closed domain. This starts from the Cauchy-Schwartz inequality
\begin{equation}
    \left|\int_V\Ab\cdot\Bb\,dV\right| \leq \int_V|\Ab\cdot\Bb|\,dV \leq \|\Ab\|\|\Bb\|,
    \label{eq:arnoldCS}
\end{equation}
where $\|\cdot\|^2 = \int_V|\cdot|^2\,dV$ denotes the $L^2$-norm of a vector field on $V$. Inequality \eqref{eq:arnoldCS} holds in any gauge, but since the helicity is gauge independent in such a domain, we are free to choose a gauge where $\Ab$ obeys a Poincar\'e inequality
\begin{equation}
    \|\Ab\| \leq C\|\Bb\|
    \label{eq:poincare}
\end{equation}
for some fixed constant $C$. For example, in a magnetically closed domain such an inequality is satisfied by the Coulomb gauge $\Ab^{\rm c}$ with $\bnabla\cdot\Ab^{\rm c}=0$ in $V$ and $\nb\cdot\Ab^{\rm c}=0$ on $\partial V$ \citep[e.g.][]{yoshida1990rot}. Then we obtain the desired energy bound
\begin{equation}
    \left|\int_V\Ab\cdot\Bb\,dV\right| \leq C\|\Bb\|^2.
    \label{eq:arnold}
\end{equation}
Next consider our magnetically-open spherical shell.

\subsection{Lower Bound for Energy} \label{sec:energy1}

For $H_{\rm R}$, we are no longer free to choose the vector potential in \eqref{eq:h}, but must verify \eqref{eq:poincare} for our particular gauge $\Ab^*$ from \eqref{eq:ptA} and \eqref{eq:tgauge}. We find that this can be done using the integral expression \eqref{eq:agreens} for $\Ab^*$ in terms of $\Bb$ alone, by adapting the argument of \citet{cantarella2001bounds}. As detailed in Appendix \ref{app:poincare}, this leads to
\begin{equation}
    \|\Ab^*\| \leq \frac{\pi r_1}{2}\|\Bb\|,
    \label{eq:poincarept}
\end{equation}
and hence \eqref{eq:arnoldCS} gives
\begin{equation}
     |H_{\rm R}| \leq \frac{\pi r_1}{2}\|\Bb\|^2,
\end{equation}
or equivalently the energy bound
\begin{equation}
    \|\Bb\|^2 \geq W_1 \equiv \frac{2|H_{\rm R}|}{\pi r_1}
    \label{eq:bound1}
\end{equation}
for our spherical shell.

\begin{figure*}
    \centering
    \includegraphics[width=\linewidth]{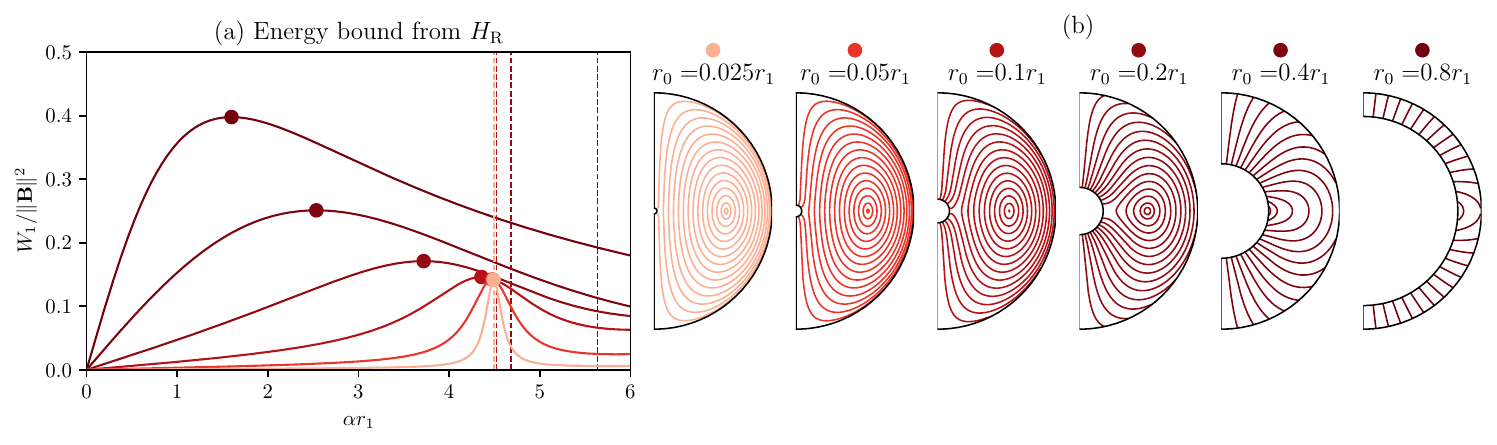}
    \caption{Linear force-free fields matching $B_r$ from \eqref{eq:pfss} on both $S_0$ and $S_1$. Panel (a) shows the ratio $W_1/\|\Bb\|^2$ in energy bound \eqref{eq:bound1} for these fields, as a function of (normalized) $\alpha$. Curves represent domains with different $r_0/r_1$, with colored dots showing the maximizing solution in each case. Panel (b) shows the flux surfaces for these maximizing solutions. Dashed vertical lines in (a) show the lowest resonant values of $\alpha$ (Appendix \ref{app:lfff}).}
    \label{fig:lfff-bounds}
\end{figure*}

The bound \eqref{eq:bound1} is not tight, in the sense that equality may be impossible to attain for any $\Bb$, at least for certain boundary conditions. To see this, note first that a magnetic field that maximizes the ratio $W_1/\|\Bb\|^2$ -- in other words, maximizes the tightness of the bound -- must be a linear force-free field satisfying
\begin{equation}
    \bnabla\times\Bb = \alpha\Bb, \quad \alpha=\textrm{constant}.
    \label{eq:lfff}
\end{equation}
This follows from Woltjer's Theorem \citep{woltjer1958theorem}, as extended by \citet{berger1984dissipation} to the case of $H_{\rm R}$ with fixed $B_r|_{\partial V}$ (we give an alternative proof in Appendix \ref{app:woltjer} using $\Ab^*$).

Then consider a simple example where we fix $B_r$ on both boundaries to match the dipolar ``source surface'' potential field,
\begin{equation}
    \Bp = B_0\left[\frac{r_0^3(r^3 + 2r_1^3)}{r^3(r_0^3 + 2r_1^3)}\cos\theta\ebr - \frac{r_0^3(r^3 - r_1^3)}{r^3(r_0^3 + 2r_1^3)}\sin\theta\ebth\right].
    \label{eq:pfss}
\end{equation}
(This is the unique curl-free field with $B_r=B_0\cos\theta$ on $S_0$ that is purely radial on $S_1$.) There is a whole family of linear force-free fields with $B_r$ matching \eqref{eq:pfss} on both boundaries, having different values of $\alpha$ (details in Appendix \ref{app:lfff}), and different values of the ratio $W_1/\|\Bb\|^2$. This ratio is plotted in Figure \ref{fig:lfff-bounds}(a) as a function of (non-dimensionalized) $\alpha r_1$, with each curve corresponding to a different ratio $r_0/r_1$ for the domain. For any $\alpha r_1$, the ratio $W_1/\|\Bb\|^2$ for these linear force-free fields depends only on the ``shape'' $r_0/r_1$ of the domain, and not the absolute ``size'' $r_1$. The dots indicate the solution that maximizes $W_1/\|\Bb\|^2$ for each domain shape, and the field line topology of each of these solutions is indicated in Figure \ref{fig:lfff-bounds}(b). Here the contours show surfaces of constant $rR(r)\sin^2\theta$, where $R(r)$ is defined in Appendix \ref{app:lfff}. These are flux surfaces (tangent to $\Bb$) in these axisymmetric fields. Notice that the topology of the maximal-helicity solution changes depending on the domain: it is a sheared arcade for $r_0=0.8r_1$, but contains a disconnected ``flux rope'' when $r_0$ is small enough. 

Returning to the tightness of the bound \eqref{eq:bound1}, we see from Figure \ref{fig:lfff-bounds}(a) that it is tightest -- meaning largest ratio $W_1/\|\Bb\|^2$ -- for larger $r_0/r_1$, but for the thinnest shell illustrated ($r_0=0.8r_1$), we reach only $W_1\approx 0.4\|\Bb\|^2$ for the maximal-helicity solution. Conversely, as $r_0\to 0$, the maximal ratio tends to a limit of $W_1 \approx 0.14\|\Bb\|^2$. In this limit, the corresponding $\alpha r_1$ converges to the smallest curl eigenvalue for a unit ball, which is approximately $4.4934$ \citep[][Table I]{cantarella2000curl-spectrum}.
For non-zero $r_0$, the ratio that is possible to attain will depend on the boundary conditions for $B_r$; more realistic configurations for the Sun will be considered in Section  \ref{sec:eg}.
 
\subsection{Lower Bound for Free Energy} \label{sec:bfree}

Since $H_{\rm R}=0$ for a potential field, one might suspect that having a non-zero $H_{\rm R}$ might imply not only a minimum amount of (total) magnetic energy, $\|\Bb\|^2$, as shown in \eqref{eq:bound1}, but also some minimum amount of free magnetic energy, $\|\Bj\|^2$, where
\begin{equation}
    \|\Bj\|^2 = \|\Bb-\Bp\|^2 = \|\Bb\|^2 - \|\Bp\|^2.
\end{equation}
Here we prove this for our spherical shell.

To derive a bound for $\|\Bj\|^2$ in terms of $H_{\rm R}$, we recall the decomposition
\begin{equation}
    H_{\rm R} = H_{\rm J} + H_{\rm PJ},
\end{equation}
where the two terms
\begin{equation}
     H_{\rm J} = \int_V\Aj\cdot\Bj\,dV, \quad H_{\rm PJ} = 2\int_V\Ap\cdot\Bj\,dV
     \label{eq:hj}
\end{equation}
are individually gauge invariant \citep{berger1999}. They have been called the ``current-carrying'' and ``volume-threading'' helicities, respectively \citep{linan2018}. Choosing $\Aj=\Aj^*$ allows us to immediately apply \eqref{eq:bound1} to $\Bj$ to see that
\begin{equation}
    |H_{\rm J}| \leq \frac{\pi r_1}{2}\|\Bj\|^2.
    \label{eq:boundhj}
\end{equation}
This already gives us the free energy bound
\begin{equation}
    \|\Bj\|^2 \geq W_2 \equiv \frac{2|H_{\rm J}|}{\pi r_1},
    \label{eq:bound2}
\end{equation}
but this only implies non-zero free energy in magnetic fields with non-zero $H_{\rm J}$.

To go further, we can bound $H_{\rm PJ}$ by
\begin{equation}
    |H_{\rm PJ}| \leq 2\int_V|\Ap|\,|\Bj|\,dV \leq 2\|\Ap\|\,\|\Bj\|.
    \label{eq:hpj}
\end{equation}
Choosing $\Ap=\Ap^*$ shows from \eqref{eq:poincarept} that
\begin{equation}
    \|\Ap^*\| \leq \frac{\pi r_1}{2}\|\Bp\|
\end{equation}
and so combining we obtain the inequality
\begin{equation}
    |H_{\rm R}| \leq \frac{\pi r_1}{2}\Big(\|\Bj\|^2 + 2\|\Bp\|\,\|\Bj\|\Big).
    \label{eq:bound3a}
\end{equation}
\edit{\citet{mactaggart2025bounds} has recently published a tighter inequality of the same form, in the sense that the coefficient of $\|\Bj\|^2$ is smaller -- this is derived by noting that $\Bj$ is a closed magnetic field so that the estimate \eqref{eq:boundhj} can be replaced by the classical bound \eqref{eq:arnold}.}

Note the appearance of $\|\Bp\|$ on the right-hand side of \eqref{eq:bound3a}. We could not expect to bound $|H_{\rm R}|$ in terms of $\|\Bj\|$ alone, since in configurations where $\Bj$ and $\Bp$ are interlinked, it is possible to increase $|H_{\rm R}|$ without bound by increasing the strength of $\Bp$ while keeping $\Bj$ fixed. A simple example of this is the linear combination of a dipole and a toroidal field,
\begin{equation}
    \Bb = ar_0^3\left(\frac{2\cos\theta}{r^3}\ebr + \frac{\sin\theta}{r^3}\ebth\right) + \frac{br}{r_0}\sin\theta\ebph.
\end{equation}
Here $\Bj$ is precisely the $\phi$-component, and it is straightforward to show that
\begin{equation}
    \Ab^* = \frac{br^2}{r_0}\cos\theta\ebr + \frac{ar_0^3}{r^2}\sin\theta\ebph
\end{equation}
so that
\begin{equation}
    H_{\rm R} = \frac{8\pi abr_0^4}{3}\left(\frac{r_1^2}{r_0^2} - 1\right).
\end{equation}
We see that indeed $H_{\rm R}$ can be increased arbitrarily by increasing the amplitude $a$ of the $\Bp$ component, while holding $b$ fixed so that $\Bj$ is unchanged.

By elementary calculus, we can solve \eqref{eq:bound3a} for $\|\Bj\|$ to see that
\begin{equation}
    \|\Bj\| \geq \|\Bp\|\left(\sqrt{1 + \frac{2|H_{\rm R}|}{\pi r_1\|\Bp\|^2}} - 1\right),
    \label{eq:boundbj}
\end{equation}
and squaring gives the desired bound
\begin{equation}
    \|\Bj\|^2 \geq W_3
    \label{eq:bound3b}
\end{equation}
for
\begin{equation}
    W_3 \equiv \frac{2|H_{\rm R}|}{\pi r_1} + 2\|\Bp\|^2\left(1 - \sqrt{1 + \frac{2|H_{\rm R}|}{\pi r_1\|\Bp\|^2}} \right).
    \label{eq:w3}
\end{equation}
Observe that if $\Bp$ vanishes so that $\Bb\equiv \Bj$, then $H_{\rm R}\equiv H_{\rm J}$ and \eqref{eq:bound3b} reduces to \eqref{eq:bound2}. Also, $W_3$ is always non-negative: this can be seen from \eqref{eq:boundbj}.

\begin{figure*}
    \centering
    \includegraphics[width=\linewidth]{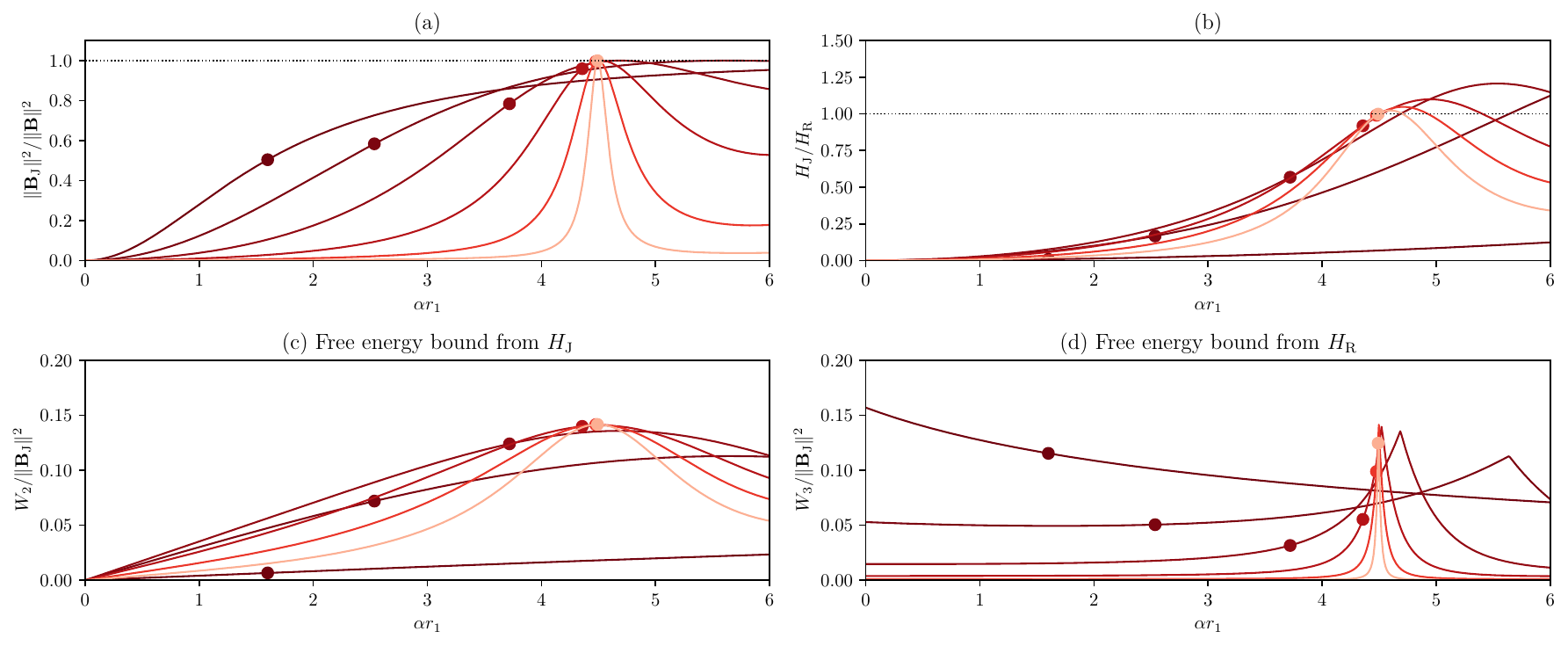}
    \caption{Normalized (a) free energy $\|\Bj\|^2$ and (b) current-carrying helicity $H_{\rm J}$, for the linear force-free solutions in Figure \ref{fig:lfff-bounds}. Panels (c) and (d) show the ratios $W_2/\|\Bj\|^2$ and $W_3/\|\Bj\|^2$ in energy bounds \eqref{eq:bound2} and \eqref{eq:bound3b}, respectively. Curves represent the same domains ($r_0/r_1$) as Figure \ref{fig:lfff-bounds}, and colored dots all show the fields maximizing the original bound $W_1/\|\Bb\|^2$.}
    \label{fig:lfff-free-bounds}
\end{figure*}

Figure \ref{fig:lfff-free-bounds} explores the tightness of the bounds \eqref{eq:bound2} and \eqref{eq:bound3b} for the same dipolar linear force-free configurations as in Figure \ref{fig:lfff-bounds}. The dots in this figure again correspond to the maximizers of $|H_{\rm R}|/\|\Bb\|^2$. Figure \ref{fig:lfff-free-bounds}(a) shows that, for these maximizers, a greater proportion of $\|\Bb\|^2$ comes from $\|\Bj\|^2$ as $r_0\to 0$. Indeed, for the cases with $r_0 \leq 0.1r_1$, $\|\Bp\|^2$ is negligible. This is seen indirectly by how the flux surfaces in Figure \ref{fig:lfff-bounds} become ``almost closed'' and barely affected by the open boundary conditions for small $r_0$. Correspondingly, as $r_0$ decreases, more and more of $H_{\rm R}$ comes from $H_{\rm J}$, as shown in Figure \ref{fig:lfff-free-bounds}(b). As a result, as $r_0\to 0$, the tightness of the bound \eqref{eq:bound2} in Figure \ref{fig:lfff-free-bounds}(c), becomes equivalent to that of the original bound \eqref{eq:bound1} in Figure \ref{fig:lfff-bounds}(b), with $W_2\approx 0.14\|\Bj\|^2$. On the other hand, for the thinnest shell $r_0=0.8r_1$, we see from Figure \ref{fig:lfff-free-bounds}(b) that $H_{\rm R}\approx H_{\rm PJ}$, so that $W_2/\|\Bj\|^2\to 0$ and the bound \eqref{eq:bound2} becomes useless, even though $\|\Bj\|^2$ is still about half of $\|\Bb\|^2$.

By contrast, the tightness of the energy bound \eqref{eq:bound3b} for the linear force-free maximizers shows a non-monotonic dependence on $\alpha$ (Figure \ref{fig:lfff-free-bounds}d). Notably, it is bounded away from zero for all $r_0$, so that non-zero free energy is guaranteed whether $H_{\rm R}$ is coming predominantly from $H_{\rm J}$ or $H_{\rm PJ}$, or from a mixture. On the other hand, for small $r_0$ the $H_{\rm R}$-based bound $W_3$ does not become independent of $r_0$, and remains slightly lower than $W_2$ at $r_0=0.025r_1$, even though $\|\Bp\|^2$ is negligible compared to $\|\Bj\|^2$.

\section{Local Energy Bounds} \label{sec:decomp}

The bounds derived in Section \ref{sec:energy} have been based on the global $H_{\rm R}$ value, integrated over the whole of $V$. Here, we spatially decompose $H_{\rm R}$ into a sum over more localized invariants, leading to stronger energy bounds. It would, of course, be possible to bound $\|\Bb\|^2$ based on $\int_V|\Ab^*\cdot\Bb|\,dV$ directly, bypassing the first inequality in \eqref{eq:arnoldCS}. However, this integral is not an ideal invariant, and we prefer to bound the energy in terms of a conserved quantity that is likely to be more robust in simulation or measurement. To achieve this, we need to decompose $H_{\rm R}$ not into arbitrary volumes \citep[cf.][]{valori2020additivity}, but into subdomains separated by magnetic surfaces.

\subsection{Partitioning the Domain}

\begin{figure}
    \centering
    \includegraphics[width=\linewidth]{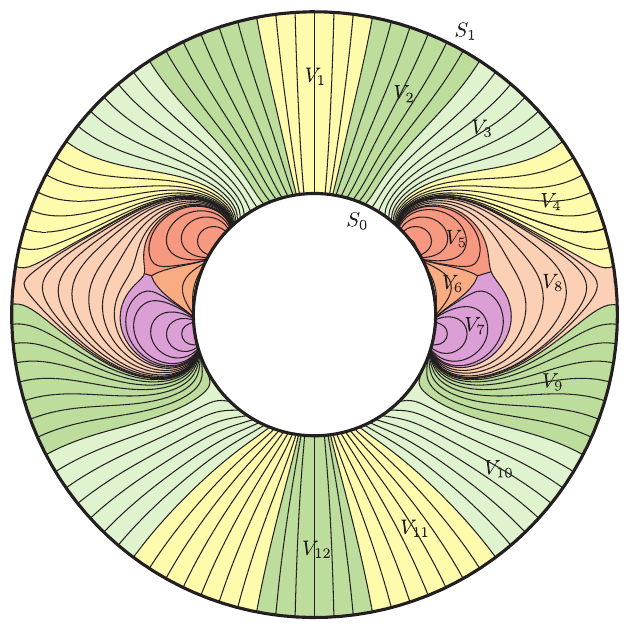}
    \caption{An example partition $\Pi$ of the spherical shell $V$ into magnetic subvolumes $V_1, \ldots, V_{12}$, in this case for an axisymmetric magnetic field. Each subvolume $V_i$ has an associated sub-helicity, $h_i$, with $\sum_{i=1}^{12}h_i = H_{\rm R}$.}
    \label{fig:decomp}
\end{figure}

The expression \eqref{EQ:HPT}
 suggests -- unlike the Finn-Antonsen formula \eqref{eq:hrfinn} -- how to spatially decompose $H_{\rm R}$ into a sum of more localized ``sub-helicities'',
\begin{equation}
    H_{\rm R} = \sum_{V_i\in\Pi}h_i,  \textrm{ where }  h_i=\int_{V_i}\Ab^*\cdot\Bb\,dV,
    \label{eq:hpart}
\end{equation}
where $\Pi = \{V_i\}$ is a disjoint partition of $V$, meaning $V=\bigcup_i V_i$ and $V_i\cap V_j=0$ for $i\neq j$. For an arbitrary such partition, each $h_i$ may not be an ideal invariant. But if the $V_i$ are magnetic subvolumes -- meaning that $\Bb\cdot\nb_i=0$ on the part of $\partial V_i$ that is internal to the overall $V$ -- then the Reynolds Transport Theorem \citep[e.g.][]{moffatt2019book}, with the uncurled ideal induction equation
\begin{equation}
    \ddy{\Ab^*}{t} = \vb\times\Bb + \bnabla\phi^*,
\end{equation}
gives
\begin{equation}
    \frac{dh_i}{dt} = \oint_{\partial V_i\cap\partial V}\big(\phi^* + \Ab^*\cdot\vb\big)\Bb\cdot\,\boldsymbol{dS}.
\end{equation}
With the global line-tied boundary condition $\vb|_{\partial V}=\bfzero$, the second term in the parentheses vanishes, and it follows also that $\bnabla_h\phi^*|_{\partial V}=0$. This means that we can take $\phi^*$ outside the integral on each boundary surface $S_0$ and $S_1$, so by \eqref{eq:nonet} we have
\begin{equation}
    \frac{dh_i}{dt} = 0.
\end{equation}
In other words, each sub-helicity is invariant under ideal motions that vanish on the global boundaries. Thus we have decomposed $H_{\rm R}$ into more localized invariants, albeit that the computation of $\Ab^*$ is defined globally. An example of such a partition is shown in Figure \ref{fig:decomp}.

\subsection{Bound from a Magnetic Partition} \label{sec:partition}

We can exploit the partition \eqref{eq:hpart} to give a stronger lower bound on the magnetic energy. To do this, define the unsigned helicity of the partition as
\begin{equation}
    \overline{H}_\Pi = \sum_{V_i\in\Pi}|h_i|.
    \label{eq:hbar}
\end{equation}
This is, again, an ideal invariant provided the $V_i$ are magnetic subvolumes. It satisfies
\begin{align}
    \overline{H}_\Pi &= \sum_{V_i\in\Pi}\left|\int_{V_i}\Ab^*\cdot\Bb\,dV\right|\\
    &\leq \sum_{V_i\in\Pi}\int_{V_i}|\Ab^*\cdot\Bb|\,dV = \int_V|\Ab^*\cdot\Bb|\,dV.
\end{align}
This recovers \eqref{eq:arnoldCS}, so it follows from Section \ref{sec:energy} that  $\overline{H}_\Pi$ for any disjoint partition $\Pi$ also obeys
\begin{equation}
    \overline{H}_\Pi \leq \frac{\pi r_1}{2}\|\Bb\|^2.
    \label{eq:bound4}
\end{equation}
For the Sun's corona, where there are positive and negative regions of helicity, a fine enough choice of partition will typically give a larger lower bound on the energy than \eqref{eq:bound1}. 

\subsection{Bound From Field Line Helicity} \label{sec:flh}

To obtain the tightest possible bound for $\|\Bb\|^2$ using \eqref{eq:bound4}, we seek the finest possible partition $\Pi$. This is given by taking each $V_i$ to be a thin flux tube around a single magnetic field line $L_i$, and taking the limit that these tube radii $r(V_i)$ go to zero. Since $h_i$ would vanish in this limit, we normalize by the tube's magnetic flux $\Phi_i$ to obtain a finite value
\begin{equation}
    \mathcal{A}(L_i) = \lim_{r(V_i)\to 0}\frac{1}{\Phi_i}\int_{V_i}\Ab^*\cdot\Bb\,dV.
    \label{eq:flhlim}
\end{equation}
Called field line helicity \citep{yeates2024flhreview}, it was shown by \citet{berger1988flh} showed that this quantity has the simpler expression
\begin{equation}
    \mathcal{A}(L_i) = \int_{L_i}\Ab^*\cdot\dl.
    \label{eq:flhline}
\end{equation}
Again, this is invariant for every field line under a line-tied ideal evolution. For a closed loop field line, $\mathcal{A}(L_i)$ is simply the enclosed magnetic flux. For an open field line, it still corresponds to a flux, although the interpretation is more nuanced \citep[e.g.,][]{yeates2016global-helicity}. We recover $H_{\rm R}$ by replacing the sum in \eqref{eq:hpart} by an integral of $\mathcal{A}(L_i)d\Phi_i$ over all field lines. In the (typical) case where all field lines intersect $\partial V$ exactly twice, this may be written as
\begin{equation}
    H_{\rm R} = \frac12\oint_{\partial V}\mathcal{A}|B_r|\,dS,
    \label{eq:hrflh}
\end{equation}
where $\mathcal{A}$ is integrated along the field line traced from each point on $\partial V$.

The corresponding unsigned helicity of this limiting ``field line'' partition is similarly given by replacing the sum in \eqref{eq:hbar} with an integral, so that
\begin{equation}
    \overline{H} = \frac12\oint_{\partial V}|\mathcal{A}B_r|\,dS.
    \label{eq:hbar2}
\end{equation}
This was called ``unsigned helicity'' by \citet{yeates2020minimal-helicity}. From \eqref{eq:bound4}  this again obeys the inequality
\begin{equation}
    \overline{H} \leq \frac{\pi r_1}{2}\|\Bb\|^2.
\end{equation}
Since this is the finest possible (magnetic) partition, it gives us the tightest possible energy bound from the partition approach, namely
\begin{equation}
    \|\Bb\|^2 \geq W_4 \equiv \frac{2\overline{H}}{\pi r_1}.
    \label{eq:bound5}
\end{equation}

It is clear that \eqref{eq:bound5} will reduce to the original bound \eqref{eq:bound1} precisely when $\overline{H}=H_{\rm R}$. This occurs when $\mathcal{A}$ either vanishes or has uniform sign for all field lines. It is not so common for $\mathcal{A}$ to vanish uniformly: as shown by \citet{yeates2020minimal-helicity}, even (non-axisymmetric) potential fields typically have non-zero $\mathcal{A}$ values on individual field lines. But the second case -- uniform sign of $\mathcal{A}$ -- can occur, at least in artificial configurations. An example is the family of linear force-free fields from Section \ref{sec:energy} (Appendix \ref{app:lfff}). Bound \eqref{eq:bound5} therefore gives no new information for these fields. On the other hand, the more typical situation in the solar corona is to have mixed signs of $\mathcal{A}$, in which case $\overline{H}> H_{\rm R}$ \citep[e.g.,][]{yeates2024cycle24} and bound \eqref{eq:bound5} constrains $\|\Bb\|^2$ more strongly. This will be illustrated in the following section.

\section{Numerical Example} \label{sec:eg}

It is clear from observations that the real solar corona does not resemble the maximal helicity fields illustrated in Figure \ref{fig:lfff-bounds} \citep{pevtsov2003,pevtsov2014review}. In particular, there are sub-regions of both positive and negative (field line) helicity, as well as regions -- particularly in the open field -- where the field lines are likely close to potential \citep{yeates2016global-helicity}. To capture these general features, we illustrate here some recent magneto-frictional simulations of Solar Cycle 24 by \citet{yeates2024cycle24}, although our interest here is purely to get a feeling for what to expect more generally from solar coronal magnetic field models.

\subsection{Solar Cycle Simulation} \label{sec:sim}

\begin{figure*}
    \centering
    \includegraphics[width=\textwidth]{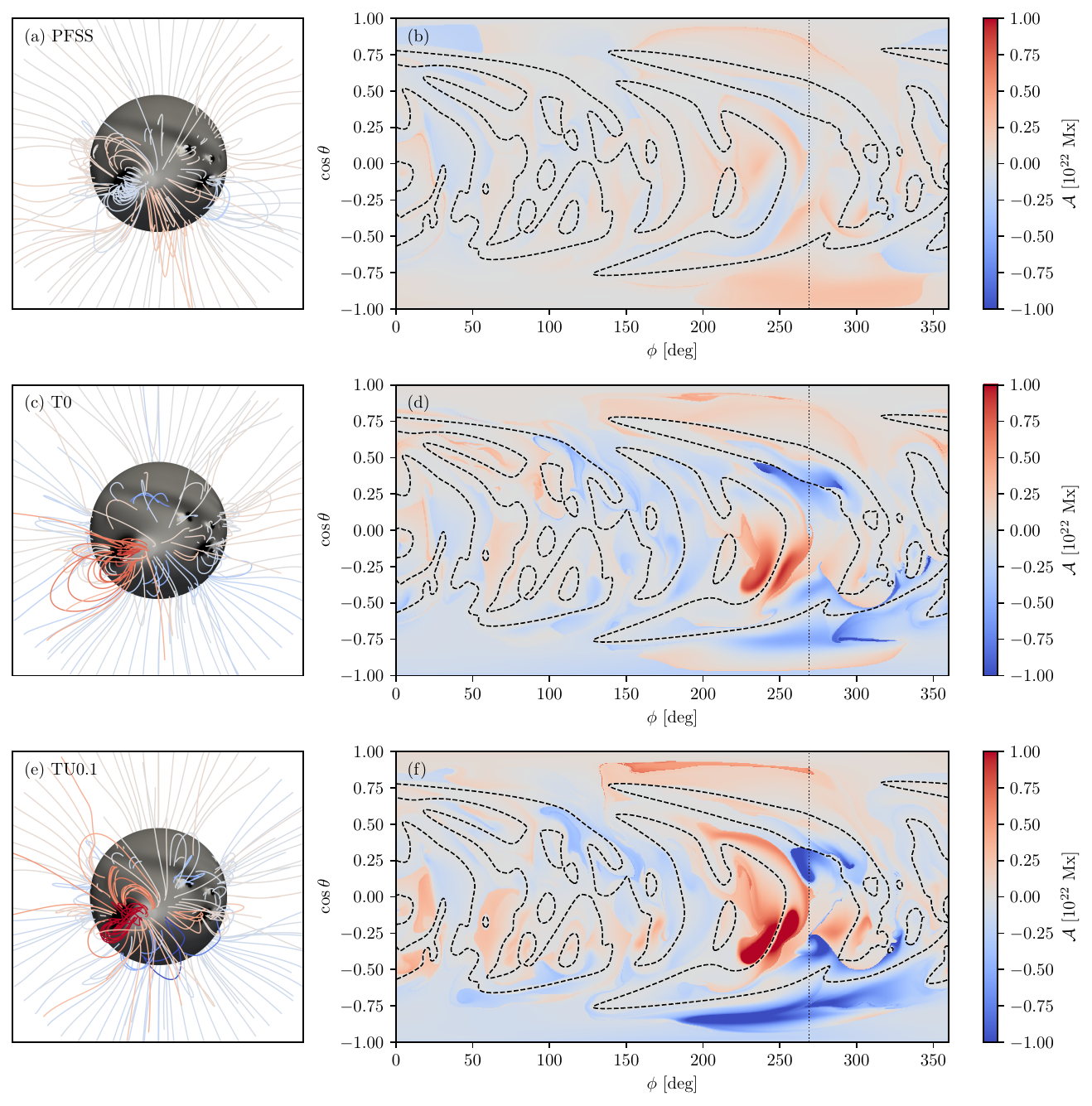}
    \caption{Field line helicity for the same day (2014 December 15) in a PFSS extrapolation (a, b), and two magneto-frictional simulations: T0 (c, d) and TU0.1 (e, f) from \citet{yeates2024cycle24}. The right column shows $\mathcal{A}$ as a function of footpoint location on $r=r_0$, with the photospheric neutral line shown dashed. The dotted vertical line shows the longitude of the viewpoint in the left column, where (selected) field lines are shown colored by $\mathcal{A}$. The color scale is capped at the same value in all plots; actual maxima are $0.4\times 10^{22}\,\mathrm{Mx}$, $1.3\times 10^{22}\,\mathrm{Mx}$ and $3.3\times 10^{22}\,\mathrm{Mx}$.}
    \label{fig:snapshots}
\end{figure*}

In the numerical simulations by \citet{yeates2024cycle24}, the lower boundary, $r_0=R_\odot$, evolves through a surface flux transport model, and the coronal magnetic field (in $r<r_1=2.5R_\odot$) responds by continual relaxation. New active regions are emerged according to SHARP (Spaceweather HMI Active Region Patch) data from Solar Dynamics Observatory \citep{bobra2014sharps}.

Figure \ref{fig:snapshots} compares two of these simulations with a potential field source surface (PFSS) extrapolation, at a single time near solar maximum. The left column shows magnetic field lines, and the right column shows the distribution of field line helicities on the solar surface (see Appendix \ref{app:num} for computational details). As detailed in \cite{yeates2024cycle24}, the two simulations have the same $B_r$ on the photosphere, but differ in the helicity of the emerging active regions. In run T0, all regions emerge untwisted, so that the helicity is injected primarily through shearing by differential rotation. In run TU0.1, the regions emerge with non-zero helicity that also contributes to the overall helicity in the corona. In both cases, the strongest helicity tends to concentrate in sheared arcades or flux ropes above neutral lines, in what would observationally correspond to filament channels \citep{martin1998}. In Figures \ref{fig:snapshots}(d,f), we see the footpoints of such structures on either side of the neutral lines. Note that, despite having $H_{\rm R}=0$, the potential field also contains non-zero field line helicity, as shown in Figure \ref{fig:snapshots}(b). This arises from the non-trivial topology of the boundary flux distribution \citep[see][]{yeates2020minimal-helicity}.

\subsection{Timeseries} \label{sec:simgraphs}

\begin{figure*}
    \centering
    \includegraphics[width=\textwidth]{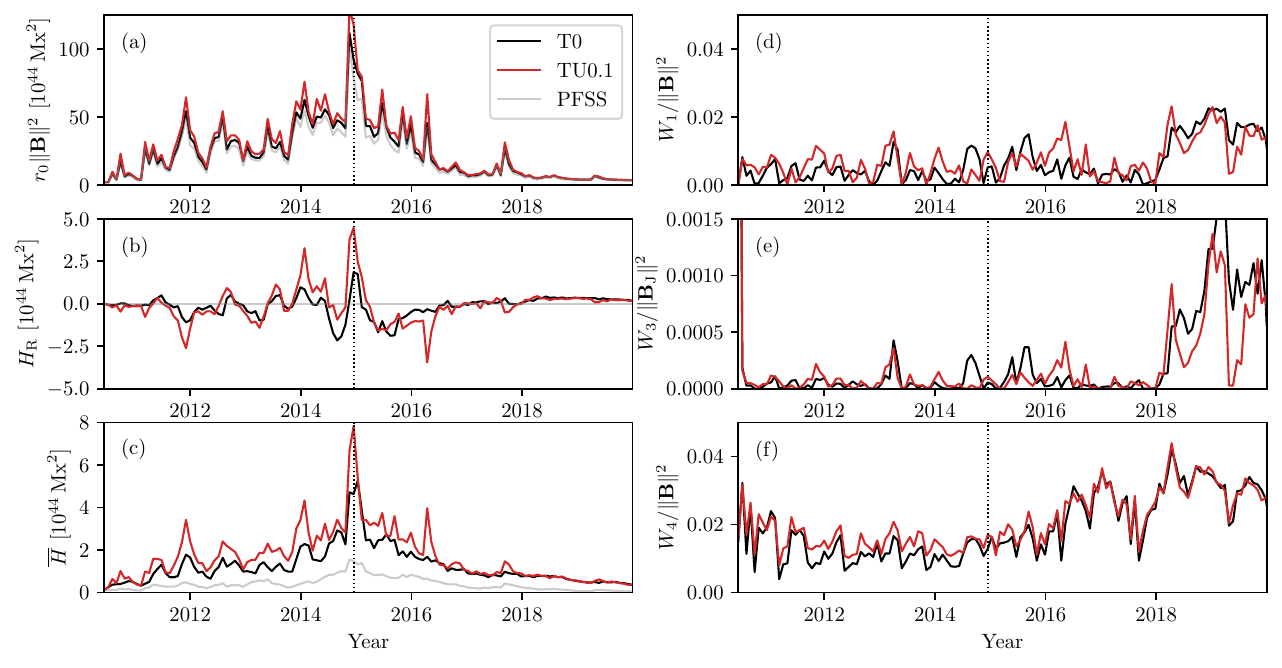}
    \caption{Time series of integrated quantities for the PFSS simulation and two magneto-frictional simulations T0, TU0.1 from \citet{yeates2024cycle24}. Left-hand panels show (a) energy $r_0\|\Bb\|^2$, (b) relative helicity $H_{\rm R}$, and (c) unsigned helicity $\overline{H}$ from equation \eqref{eq:hbar}. Right-hand panels (d-f) show the ratios $W_1/\|\Bb\|^2$, $W_3/\|\Bj\|^2$ and $W_4/\|\Bb\|^2$ in energy bounds \eqref{eq:bound1}, \eqref{eq:bound3b} and \eqref{eq:bound5}, respectively. The dotted vertical line indicates the time shown in Figure \ref{fig:snapshots}.}
    \label{fig:timeseries}
\end{figure*}

For quantitative comparison, we compute $\|\Bb\|^2$, $\|\Bj\|^2$, $H_{\rm R}$, and $\overline{H}$ from snapshots taken at 27-day intervals in the simulations T0 and TU0.1. Figure \ref{fig:timeseries}(a) shows the time variation of $\|\Bb\|^2$  (multiplied by $r_0$ to give the same units as helicity) in these two simulations as well as a sequence of corresponding PFSS extrapolations from the (shared) lower boundary distribution. Whilst the PFSS extrapolation is not precisely equal to either of the reference fields $\Bp$, owing to the different outer boundary condition, its energy is nevertheless a close approximation. Therefore this figure shows that $\|\Bp\|^2$ dominates $\|\Bj\|^2$ throughout these simulations, which is typical for many global coronal models \citep{yeates2018issi}.

Figure \ref{fig:timeseries}(b) shows how $H_{\rm R}$ varies between positive and negative, depending on the dominant non-potential structure(s). For example, at the time of the snapshot in Figure \ref{fig:snapshots}, we have $H_{\rm R}>0$ in both T0 and TU0.1, corresponding to dominance of the large active region around $\cos\theta=-0.25$ and $\phi=250^\circ$. Although this does not appear to dominate in Figures \ref{fig:snapshots}(e) and (f), it does dominate once the flux weighting $|B_r|$ is included in \eqref{eq:hrflh}. By definition, $H_{\rm R}=0$ for the PFSS model, since it is a current-free field. By contrast, the unsigned helicity in Figure \ref{fig:snapshots}(c) is always non-negative, shows a more consistent ordering with the value for TU0.1 being larger than for T0, and has a non-zero value (though smaller) for the PFSS model.

\subsection{Energy Bounds} \label{sec:simbounds}

The right column of Figure \ref{fig:timeseries} illustrates the tightness of three of the energy bounds from Sections \ref{sec:energy} and \ref{sec:decomp} over the simulations T0 and TU0.1. For the basic bound \eqref{eq:bound1} on $\|\Bb\|^2$ using $|H_{\rm R}|$, shown in Figure \ref{fig:timeseries}(a), non-zero bounds are obtained whenever $H_{\rm R}\neq 0$, but note that the ratio $W_1/\|\Bb\|^2$ is always substantially smaller than for the linear force-free example in Figure \ref{fig:lfff-bounds}. This arises in part from the fact that $H_{\rm R}$ now incorporates cancelling contributions of positive and negative sign, and in part from the fact that the numerical models are not linear force-free fields and so not maximizers of the ratio. There is a noticeable increase in the ratio after 2018 when $H_{\rm R}$ becomes uniformly positive, owing to dominance of a particular long-lived structure.

Figure \ref{fig:timeseries}(e) shows the ratio $W_3/\|\Bj\|^2$, namely the tightness of the bound \eqref{eq:bound3b} on free energy $\|\Bj\|^2$ using $H_{\rm R}$ and $\|\Bp\|^2$. Whilst this is again non-zero at times where $H_{\rm R}\neq 0$, the tightness ratio is orders of magnitude lower than the first bound. To understand this behavior of bound \eqref{eq:bound3b}, divide \eqref{eq:w3} through by $\|\Bp\|^2$ and set $\varepsilon \equiv W_1/\|\Bp\|^2$. From Figure \ref{fig:timeseries}(d), and the fact that $\|\Bj\|^2<\|\Bp\|^2$, we see that $\varepsilon\ll 1$. Then use Taylor expansion around $\varepsilon=0$ to see that $W_3 \approx \varepsilon^2\|\Bp\|^2/4$, so that $W_3/\|\Bj\|^2 \ll \varepsilon$ provided that $\|\Bj\|^2$ is not too small. So the lack of tightness of the free energy bound would seem to be quite a generic situation for global coronal models.

Finally, Figure \ref{fig:timeseries}(f) shows the tightness of the improved energy bound  \eqref{eq:bound5} on $\|\Bb\|^2$ based on the unsigned helicity. Overall the ratio $W_4/\|\Bb\|^2$ is larger than $W_1/\|\Bb\|^2$, although never exceeding 0.05 (only a fifth of that reached for the maximal helicity linear force-free field in the dipole example with $r_0=0.4r_1$). Most significantly, perhaps, this unsigned helicity bound is always non-zero in both runs T0 and TU0.1, unlike the bounds based on $H_{\rm R}$. This illustrates how $\overline{H}$ may be better suited than $H_{\rm R}$ to analyzing these global simulations.

\section{Conclusion} \label{sec:conclusion}

The bounds derived above show that non-zero relative helicity $|H_{\rm R}|$ implies non-zero magnetic energy, analogous to classical helicity $|H|$ in a magnetically-closed domain. Similarly, we have shown that non-zero $|H_{\rm R}|$ implies non-zero \emph{free} magnetic energy. However, just as with $H$, magnetic fields can put non-trivial topological constraints on the energy even when $H_{\rm R}=0$. The unsigned helicity $\overline{H}$ in \eqref{eq:hbar2} seems the most natural generalization, and we have shown that if this is non-zero then it implies non-zero magnetic energy even in fields with $H_{\rm R}=0$. 

In the classical magnetically-closed case, the $\Bb$ that maximizes helicity for a given magnetic energy depends only on the geometry of the domain \citep{arnold1986hopf,cantarella2001bounds}. In the case of $H_{\rm R}$, the helicity that can be reached 
-- and nature of the maximizing configuration -- depend also on the boundary distribution of $\nb\cdot\Bb$. We illustrated the maximizing $\Bb$ for the simplest case of a dipole boundary condition, but in future it could be computed (with care) for arbitrary distributions, including those from the numerical simulations in Section \ref{sec:eg}. This could provide a normalized way to assess how ``helical'' is a given magnetic configuration in the solar corona, with possible application to predicting solar eruptions.

What about domains other than spherical shells? Firstly, we expect all of our results to extend (with small modification) to the case of an infinite plane slab (the volume between two parallel planes), since the poloidal-toroidal decomposition of $\Bb$ also exists there. We omit this here since such a domain is of limited practical relevance in solar physics. More relevant would be a finite domain, such as a Cartesian box. For such a domain, \citet{demoulin2007review} notes that we can again write $H_{\rm R}=\int_V\Ab^{\rm c}\cdot\Bb\,dV$, by choosing $\Ab^{\rm c}$ to match $\nb\times\Ab^{\rm c}=\nb\times\Ap^{\rm c}$ on $\partial V$, where the reference vector potential $\Ap^{\rm c}$ satisfies the Coulomb gauge conditions $\nabla\cdot\Ap^{\rm c}=0$ throughout $V$ and $\nb\cdot\Ap^{\rm c}|_{\partial V}=0$. However, except for spherical shells or infinite slabs, this does not yield an explicit expression for $\Ab^{\rm c}$ in terms of $\Bb$, so that it is unclear how to derive an energy bound \edit{for $\|\Bb\|^2$, although \citet{mactaggart2025bounds} has recently derived a free energy bound similar to Section \ref{sec:bfree}.}

In certain other cases -- for example, a doubly-periodic domain \citep{xiao2025periodic}, or a finite flux tube \citep{prior2014winding-gauge, candelaresi2021self-mutualwinding} -- there are known $\Ab$ with explicit expressions in terms of $\Bb$ and associated interpretations of the corresponding $H$ in terms of winding numbers. However, in these cases, $H$ can differ from $H_{\rm R}$ \citep[an explicit example is given by][]{prior2014winding-gauge}. This raises the question whether $H_{\rm R}$ is the most physically relevant helicity definition in general, or whether it coincides with such a definition only in more specialized domains such as spherical shells. Indeed, \citet{mactaggart2023relativemultiply-connected} have proposed how to extend $H_{\rm R}$ to multiply-connected domains, but the relation to energy warrants further investigation.


\vspace{1cm}
ARY was supported by UKRI/STFC grant UKRI1216. The authors thank Christopher Prior for comments on an initial draft, as well as  Daining Xiao and Mitchell Berger for fruitful discussions. The SDO data are courtesy of NASA and the SDO/HMI science team.

\facility{SDO (HMI)}

%




\appendix

\section{Proof of Poincar\'e Inequality} \label{app:poincare}

To derive \eqref{eq:poincarept} for $\Ap^*$, we adapt an argument of \citet{cantarella2001bounds} which was applied originally to the Biot-Savart integral in a magnetically closed domain. We start from the expression \eqref{eq:agreens} for $\Ab^*$, using \eqref{eq:gradg} to write it in the form
\begin{equation}
    \Ab^*(\xb) = \frac{1}{4\pi r^3}\int_{S_r'}\left[B_r(\xb')\frac{\xb'\times\xb}{1-\cos\xi} + \frac{\Bb(\xb')\cdot(\xb\times\xb')}{1-\cos\xi}\ebr\right]\,dS'.
\end{equation}
Since $\xb=r\ebr$, the two terms in the integrand are orthogonal vectors for any $\xb'$, so
\begin{equation}
    |\Ab^*(\xb)|^2 = \frac{1}{(4\pi r^3)^2}\left[\left|\int_{S_r'}B_r(\xb')\frac{\xb'\times\xb}{1-\cos\xi}\,dS'\right|^2 + \left|\int_{S_r'}\frac{\Bb(\xb')\cdot(\xb\times\xb')}{1-\cos\xi}\,dS'\right|^2 \right].
    \label{eq:b6}
\end{equation}
The first term in \eqref{eq:b6} satisfies
\begin{align}
\left|\int_{S_r'}B_r(\xb')\frac{\xb'\times\xb}{1-\cos\xi}\,dS'\right| &\leq \int_{S_r'}\left||B_r(\xb')|\frac{|\xb'\times\xb|^{1/2}}{(1-\cos\xi)^{1/2}}\right|\frac{|\xb'\times\xb|^{1/2}}{(1-\cos\xi)^{1/2}}\,dS'\\
&\leq r^2\left(\int_{S_r'}|B_r(\xb')|^2\Gamma(\xi)\,dS'\right)^{1/2}\left(\int_{S_r'}\Gamma(\xi)\,dS'\right)^{1/2}.
\end{align}
using Cauchy-Schwartz and the fact that $|\xb'\times\xb|=r^2\sin\xi$. As in \eqref{eq:hwinding}, we use the shorthand $\Gamma(\xi)=\sin\xi/(1-\cos\xi)$. Similarly the second term in \eqref{eq:b6} satisfies
\begin{align}
    \left|\int_{S_r'}\frac{\Bb(\xb')\cdot(\xb\times\xb')}{1-\cos\xi}\,dS'\right| \leq r^2\left(\int_{S_r'}|\Bb_h(\xb')|^2\Gamma(\xi)\,dS'\right)^{1/2}\left(\int_{S_r'}\Gamma(\xi)\,dS'\right)^{1/2}.
\end{align}
By symmetry, the integral $\int_{S_r}\Gamma(\xi)\,dS'$ must be independent of $\xb$, so can be directly evaluated by setting $\xb$ to be the north pole, giving
\begin{equation}
    \int_{S_r'}\Gamma(\xi)\,dS' = 2\pi^2 r^2.
\end{equation}
So overall \eqref{eq:b6} gives
\begin{align}
    |\Ab^*(\xb)|^2 &\leq \frac{1}{8}\int_{S_r'}|\Bb(\xb')|^2\Gamma(\xi)\,dS'.
\end{align}
Integrating over all $\xb$ in $S_r$ then gives
\begin{equation}
    \int_{S_r}|\Ab^*(\xb)|^2\,dS \leq \frac{1}{8}\int_{S_r'}|\Bb(\xb')|^2\left(\int_{S_r}\Gamma(\xi)\,dS\right)\,dS' = \frac{\pi^2 r^2}{4}\int_{S_r'}|\Bb(\xb')|^2\,dS'.
\end{equation}
Finally, integrating over $r$ gives
\begin{equation}
    \|\Ab^*(\xb)\|^2 \leq \int_{r_0}^{r_1}\frac{\pi^2r^2}{4}\int_{S_r}|\Bb(\xb')|^2\,dS'\,dr \leq \frac{\pi^2r_1^2}{4}\|\Bb(\xb)\|^2.
\end{equation}

\section{Woltjer's Theorem for Relative Helicity} \label{app:woltjer}

\citet{berger1984dissipation} (in Appendix A) showed that the celebrated theorem of \citet{woltjer1958theorem} applies equally to relative helicity: namely, the minimum energy state (if it exists) for a specified $H_{\rm R}$ and  $\Bb\cdot\nb|_{\partial V}$ must be a linear force-free field satisfying \eqref{eq:lfff}.
Here we use \eqref{EQ:HPT} to give a simple alternative proof for our spherical shell domain.

Start by considering the first variation of $\|\Bb\|^2$. Integration by parts gives
\begin{equation}
    \delta\|\Bb\|^2 = 2\int_V\delta\Ab^*\cdot(\bnabla\times\Bb)\,dV - 2\oint_{\partial V}\Bb\cdot(\delta\Ab^*\times\nb)\,dS.
\end{equation}
The surface term vanishes since $\delta\Ab^*\times\nb=\bfzero$, because $\Ab^*\times\nb|_{\partial V}$ in the poloidal-toroidal gauge \eqref{eq:ptA} depends only on $B_r|_{\partial V}$, which is fixed by assumption. On the other hand, and applying a similar boundary condition, we find
\begin{equation}
    \delta H_{\rm R} = 2\int_V\delta\Ab^*\cdot\Bb\,dV.
\end{equation}
We then follow the original argument of \citet{woltjer1958theorem} to minimize $\|\Bb\|^2$ with fixed $H_{\rm R}$ and fixed $B_r$ on the boundaries, introducing a Lagrange multiplier $\alpha$ to obtain
\begin{equation}
    \int_V2\delta\Ab^*\cdot(\bnabla\times\Bb - \alpha\Bb)\,dV = 0.
\end{equation}
This holds for all allowable perturbations $\delta\Ab^*$ if and only if \eqref{eq:lfff} is satisfied.

\section{Dipolar Linear Force-Free example} \label{app:lfff}

By separation of variables \citep[e.g.,][]{berger1985lfff, durrant1989lfff}, it may be shown that a solution to \eqref{eq:lfff} with $B_r$ matching \eqref{eq:pfss} on $S_0$ and $S_1$ takes the form
\begin{equation}
\Bb = B_0\left[\frac{2R\cos\theta}{r}\ebr - \frac{(rR)'\sin\theta}{r}\ebth + \alpha R\sin\theta\ebph\right],
\label{eq:lfffb}
\end{equation}
where the prime denotes differentiation with respect to $r$, and the radial function $R(r)$ is given in terms of spherical Bessel functions as
\begin{equation}
    R(r) = Aj_1(\alpha r) + By_1(\alpha r),
    \label{eq:R}
\end{equation}
with constants $A$, $B$ determined by the boundary conditions on $S_0$ and $S_1$. Specifically, 
\begin{align}
A = - \frac{r_0y_1(\alpha r_1) - cr_1y_1(\alpha r_0)}{2\lambda}, \qquad B =  \frac{r_0j_1(\alpha r_1) - cr_1j_1(\alpha r_0)}{2\lambda},
\end{align}
where we have used the shorthand
\begin{equation}
    \lambda \equiv y_1(\alpha r_0)j_1(\alpha r_1) - j_1(\alpha r_0)y_1(\alpha r_1), \qquad c \equiv \frac{3r_0^3}{r_0^3 + 2r_1^3}.
\end{equation}
This solution exists for every value of $\alpha \in\mathbb{R}$ except for the discrete set of ``resonant'' values where $\lambda=0$ \citep{berger1985lfff}. These correspond to the values of $\alpha$ for which the homogeneous problem with $B_r(r_0,\theta,\phi)=B_r(r_1,\theta,\phi)=0$ has a non-trivial solution -- the curl eigenfunctions in a spherical shell \citep{cantarella2000curl-spectrum}. The lowest of these resonant $\alpha$ values for each $r_0$ is indicated by a dashed vertical line in Figure \ref{fig:lfff-bounds}(a). 

For a linear force-free field, we have $\ebr\cdot\bnabla\times\Bb = \alpha B_r$, so since $\cos\theta$ is an eigenfunction of $\nabla_h^2$, the solutions to \eqref{eq:lapP} and \eqref{eq:lapT} are
\begin{equation}
    P(r,\theta) = B_0 r R\cos\theta, \quad T(r,\theta)=\alpha P(r,\theta),
\end{equation}
with resulting vector potential
\begin{equation}
    \Ab^* = B_0\big[\alpha rR\cos\theta \ebr + R\sin\theta\ebph\big].
\end{equation}
By \eqref{EQ:HPT}, the relative helicity \citep[cf.][]{berger1985lfff} is therefore
\begin{equation}
    H_{\rm R} = \int_V\Ab^*\cdot\Bb\,dV = \frac{16\pi B_0^2\alpha}{3}\int_{r_0}^{r_1}(rR)^2\,dr.
    \label{eq:lfffhr}
\end{equation}
Similarly the energy reduces to the single integral
\begin{equation}
    \|\Bb\|^2 = \frac{8\pi B_0^2}{3}\int_{r_0}^{r_1}\big[(2+\alpha^2r^2)R^2 + ([rR]')^2 \big]\,dr.
\end{equation}
At the resonant values of $\alpha$, both $\|\Bb\|^2$ and $H_{\rm R}$ diverge to infinity, although their ratio -- as shown in Figure \ref{fig:lfff-bounds}(a) -- does not, because the factor of $\lambda^{-2}$ cancels between them.

To compute $H_{\rm J}$ and $H_{\rm PJ}$, it is convenient to use the vector potential
\begin{equation}
    \Ap = \frac{B_0r_0^3(r^3 + 2r_1^3)}{2r^2(r_0^3 + 2r_1^3)}\sin\theta\ebph
\end{equation}
for the reference field \eqref{eq:pfss}, since in this gauge $\Ap\cdot\Bp=0$ and we have
\begin{align}
    H_{\rm PJ} = 2\int_V\Ap\cdot\Bb\,dV = \frac{8\pi B_0^2 r_0^3\alpha}{3(r_0^3 + 2r_1^3)}\int_{r_0}^{r_1}(r^3 + 2r_1^3)R\,dr,
\end{align}
from which we compute $H_{\rm J}$ using $H_{\rm J}=H_{\rm R}- H_{\rm PJ}$.

Finally, note that the field line helicity $\mathcal{A}$ in these fields has the same sign for every field line. This follows from \eqref{eq:flhlim} noting that
\begin{equation}
    \Ab^*\cdot\Bb = \alpha B_0^2R^2(1+\cos^2\theta).
\end{equation}

\section{Numerical Method for Field Line Helicity} \label{app:num}

As with all calculations of magnetic helicity, care is needed to use a numerical method that does not introduce spurious $\bnabla\cdot\Bb$ contributions \citep{valori2016issi}. Here we summarize how the calculations of $\mathcal{A}$ and hence $\overline{H}$ in Section \ref{sec:eg} were made \citep[code available at][]{sphtools-code}. As indicated by \citet{yeates2024cycle24}, the simulations use a finite-volume method and provide $\Bb$ components on the corresponding cell faces of a staggered mesh. Since neighboring cells share the face values, this ensures that $\bnabla\cdot\Bb=0$ is preserved in an integrated sense within each three-dimensional grid cell. To ensure that $\bnabla\times\Ab=\Bb$ when computing $\Ab$, the components of $\Ab$ are staggered on the cell edges, and computed in such a way as to enforce the discrete Stokes Theorem, $\int_{S_{ij}}\nb\cdot\Bb\,dS = \oint_{\partial S_{ij}}\Ab\cdot\,\dl$ on every cell face $S_{ij}$.

\subsection{Vector potential}

Rather than computing $\mathcal{A}$ directly from $\Ab^*$ using \eqref{eq:flhline}, we instead evaluate
\begin{equation}
    \mathcal{A}(L_i) = \int_{L_i}\Ab'\cdot\dl 
    \label{eq:flhprimed}
\end{equation}
where $\Ab'$ is simpler to compute. To ensure that $\mathcal{A}$ is unmodified for all field lines by this gauge transformation, we impose that (i) $\ebr\times\Ab'=\ebr\times\Ab^*$ on both $S_0$ and $S_1$, and (ii) $\oint_{S_r}A'_r\,dS = 0$ on every $S_r$ for $r\in[r_0, r_1]$. See this by writing $\Ab'=\Ab^* + \bnabla\chi$, so that condition (i) implies that $\chi(r_0,\theta,\phi)\equiv\chi_0$ and $\chi(r_1,\theta,\phi)\equiv\chi_1$, both constant. Thus
\begin{equation}
     \int_{L_i}\Ab'\cdot\dl = \mathcal{A}(L_i) + \chi_1-\chi_0.
\end{equation}
Then, noting from \eqref{eq:tgauge} that $\oint_{S_r}A^*_r\,dS = 0$, condition (ii) implies that
\begin{equation}
    \oint_{S_r}\ddy{\chi}{r}\,dS = 0 \quad \implies \frac{d}{dr}\left(\frac{1}{r^2}\oint_{S_r}\chi\,dS\right) = 0 \quad \implies \chi_1=\chi_0.
\end{equation}

\subsection{Numerical algorithm} \label{app:algo}

To compute a suitable $\Ab'$, we proceed as follows:
\begin{enumerate}
    \item Compute $\Ab^\dagger$ in the DeVore-Coulomb gauge \citep{devore2000diffrot, moraitis2018comp},
    \begin{equation}
        \Ab^\dagger(r,\theta,\phi) = \frac{r_0}{r}\bnabla\times\big[\psi(\theta,\phi)\ebr\big] + \frac{1}{r}\int_{r_0}^{r}\Bb(r',\theta,\phi)\times\ebr\,r'dr', \quad \textrm{where } \nabla_h^2\psi = -B_r(r_0,\theta,\phi),
        \label{eq:dc}
    \end{equation}
   choosing the solution $\psi$ with zero mean over $S_1$. By definition, $A^\dagger_r\equiv 0$ so $\Ab^\dagger$ already satisfies condition (ii). In addition, $\psi$ coincides with $P$ from Section \ref{sec:sph}, so $\Ab^\dagger$ also satisfies condition (i) on $S_0$, but not (necessarily) on $S_1$. Both the Poisson equation and the radial integral are discretized so that the discrete Stokes Theorem is satisfied on each cell face, with $\psi$ co-located with $B_r$ on the radial cell faces.
    \item Change gauge to $\Ab'=\Ab^\dagger + \bnabla g$ satisfying both conditions (i) and (ii), while preserving $\oint_{\partial S_{ij}}\Ab'\cdot\dl = \oint_{\partial S_{ij}}\Ab^\dagger\cdot\dl$ on every cell face. The gauge function $g$ is located at cell vertices, and determined as follows:
    \begin{enumerate}
        \item Compute the polar values $A^*_\theta(r_1,0,\phi)$ and $A^*_\theta(r_1,\pi,\phi)$ on $S_1$ by solving $\nabla^2_h P = -B_r$. These will be used as a boundary conditions in the next step.
        \item Compute the gauge function $g(r_1,\theta,\phi)$ on $S_1$ by solving $\nabla^2_h g = -\bnabla_h\cdot\Ab^\dagger$ on $S_1$, choosing the solution with zero mean (this additive constant affects only $A'_r$, not $A'_\theta$ or $A'_\phi$). This will ensure that $\bnabla_h\cdot\Ab'=0$ on $S_1$, and hence that $\ebr\times\Ab'=\ebr\times\Ab^*$ on $S_1$. Since the first and last rows of vertices are at the poles, we avoid numerical problems by solving for $g$ with these two rows omitted, setting the values of $g$ at the poles with the Neumann condition $\partial g/\partial \theta = A^*_\theta - A^\dagger_\theta$.
        \item Set $g(r_0,\theta,\phi)\equiv 0$ so as to preserve condition (i) on $S_0$.
        \item Fill in $g(r,\theta,\phi)$ for intermediate $S_r$ by linear interpolation between $g(r_0,\theta,\phi)\equiv 0$ and $g(r_1,\theta,\phi)$. This will preserve condition (ii) since $g(r_1,\theta,\phi)$ has zero mean.
    \end{enumerate}
\end{enumerate}

Once $\Ab'$ has been computed, we trace field lines $L_i$ using the (second-order) midpoint method, then evaluate \eqref{eq:flhprimed} by numerical integration of $\Ab'$ along each field line. To compute $\overline{H}$, we assume that there are no closed field lines and approximate \eqref{eq:hbar2} by tracing field lines from grids at twice the original mesh resolution on both $S_0$ and $S_1$.

\subsection{Convergence test}

\begin{figure*}
    \centering
    \includegraphics[width=\textwidth]{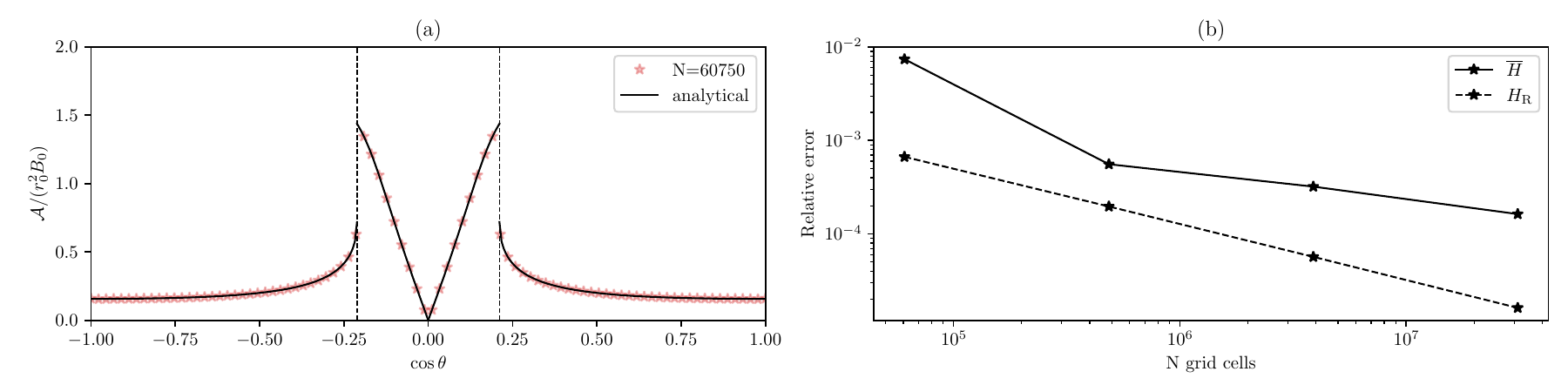}
    \caption{Numerical convergence test using the linear force-free field \eqref{eq:lfffb}. Panel (a) shows $\mathcal{A}$ as a function of footpoint colatitude on $S_0$, while panel (b) shows how the errors in numerically-approximated $\overline{H}$ and $H_{\rm R}$ depend on the mesh resolution for discretizing $\Bb$.}
    \label{fig:numerics}
\end{figure*}

Figure \ref{fig:numerics} demonstrates convergence of the numerical scheme, for a test where the 
linear force-free field \eqref{eq:lfffb} is discretized on the same three-dimensional mesh used for the simulations in Section \ref{sec:eg}. We fix $r_0=0.8$, $r_1=1$, $\alpha = 1.6$ and $B_0=1$. For \eqref{eq:lfffb}, $\mathcal{A}$ may be calculated to arbitrary precision by parameterizing \eqref{eq:flhline} with the radial component $r$, and evaluating an integral in $r$ for each field line. This ``analytical'' solution is shown by the solid curve in Figure \ref{fig:numerics}(a); note the separatrices at
\begin{equation}
    \cos\theta = \pm\sqrt{1 - \frac{r_1R(r_1)}{r_0R(r_0)}},
\end{equation}
between open field lines nearer the poles and closed field lines nearer the equator. The star symbols show the $\mathcal{A}$ profile computed entirely ``agnostically'' from the discretized datacube using the method described in Section \ref{app:algo}. In this case, the datacube resolution was only $15\times 45\times 90$ (a quarter of that used in the numerical simulations in Section \ref{sec:eg}).

The solid line in Figure \ref{fig:numerics}(b) shows the error in $\overline{H}$ when numerically computed from the datacube at resolutions from $15\times 45\times 90$ up to $120\times 360\times 720$. The ground truth for this test is given by \eqref{eq:lfffhr} since $\overline{H}=H_{\rm R}$ for \eqref{eq:lfffb}. Despite the multiple stages in the numerical calculation (discretizing $\Bb$ on the mesh, computing $\Ab'$, tracing field lines, integrating $\mathcal{A}$ along them, and integrating the results over $S_0$ and $S_1$), we observe (modest) convergence with the mesh resolution. The dashed line shows the approximation of $H_{\rm R}$ by direct numerical integration of $\Ab'\cdot\Bb$ on the three-dimensional mesh, which avoids several of these stages and is somewhat more accurate, but cannot be used to compute $\overline{H}$ when $\mathcal{A}$ has mixed signs.

\bibliography{yeates}{}

@INPROCEEDINGS{aly2014braids,
       author = {{Aly}, J. -J.},
        title = "{Lower and upper bounds on the energy of braided magnetic fields}",
    booktitle = {Journal of Physics Conference Series},
         year = 2014,
       series = {Journal of Physics Conference Series},
       volume = {544},
        month = oct,
    publisher = {IOP},
          eid = {012003},
        pages = {012003},
          doi = {10.1088/1742-6596/544/1/012003},
       adsurl = {https://ui.adsabs.harvard.edu/abs/2014JPhCS.544a2003A}
}

@ARTICLE{arnold1986hopf,
       author = {{Arnol'd}, V.~I.},
        title = "{The asymptotic Hopf invariant and its applications}",
      journal = {Sel. Math. Sov.},
         year = 1986,
       volume = {5},
        pages = {327-345}
}

@ARTICLE{backus1958-PT,
       author = {{Backus}, George},
        title = "{A class of self-sustaining dissipative spherical dynamos}",
      journal = {Annals of Physics},
         year = 1958,
        month = aug,
       volume = {4},
       number = {4},
        pages = {372-447},
          doi = {10.1016/0003-4916(58)90054-X},
       adsurl = {https://ui.adsabs.harvard.edu/abs/1958AnPhy...4..372B}
}

@ARTICLE{berger1984relative-helicity,
       author = {{Berger}, M.~A. and {Field}, G.~B.},
        title = "{The topological properties of magnetic helicity}",
      journal = {Journal of Fluid Mechanics},
         year = 1984,
        month = oct,
       volume = {147},
        pages = {133-148},
          doi = {10.1017/S0022112084002019},
       adsurl = {https://ui.adsabs.harvard.edu/abs/1984JFM...147..133B}
}

@ARTICLE{berger1984dissipation,
       author = {{Berger}, M.~A.},
        title = "{Rigorous new limits on magnetic helicity dissipation in the solar corona}",
      journal = {Geophysical and Astrophysical Fluid Dynamics},
         year = 1984,
        month = sep,
       volume = {30},
       number = {1},
        pages = {79-104},
          doi = {10.1080/03091928408210078},
       adsurl = {https://ui.adsabs.harvard.edu/abs/1984GApFD..30...79B}
}

@ARTICLE{berger1985lfff,
       author = {{Berger}, M.~A.},
        title = "{Structure and stability of constant-alpha force-free fields}",
      journal = {\apjs},
         year = 1985,
        month = nov,
       volume = {59},
        pages = {433-444},
          doi = {10.1086/191079},
       adsurl = {https://ui.adsabs.harvard.edu/abs/1985ApJS...59..433B}
}

@ARTICLE{berger1988flh,
       author = {{Berger}, M.~A.},
        title = "{An energy formula for nonlinear force-free magnetic fields}",
      journal = {\aap},
         year = 1988,
        month = aug,
       volume = {201},
       number = {2},
        pages = {355-361},
       adsurl = {https://ui.adsabs.harvard.edu/abs/1988A&A...201..355B}
}

@ARTICLE{berger1993braids,
       author = {{Berger}, M.~A.},
        title = "{Energy-crossing number relations for braided magnetic fields}",
      journal = {\prl},
         year = 1993,
        month = feb,
       volume = {70},
       number = {6},
        pages = {705-708},
          doi = {10.1103/PhysRevLett.70.705},
       adsurl = {https://ui.adsabs.harvard.edu/abs/1993PhRvL..70..705B}
}

@ARTICLE{berger1999,
       author = {{Berger}, M.~A.},
        title = "{Introduction to magnetic helicity.}",
      journal = {Plasma Physics and Controlled Fusion},
         year = 1999,
        month = dec,
       volume = {41},
        pages = {B167-B175},
          doi = {10.1088/0741-3335/41/12B/312},
       adsurl = {https://ui.adsabs.harvard.edu/abs/1999PPCF...41B.167B}
}

@INCOLLECTION{berger2003topologicalmhd,
       author = {{Berger}, Mitchell A.},
        title = "{Topological quantities in magnetohydrodynamics}",
    booktitle = {Advances in Nonlinear Dynamics},
         year = 2003,
       editor = {{Ferriz-Mas}, Antonio and {N{\'u}{\~n}ez}, Manuel},
        pages = {345-374},
          doi = {10.1201/9780203493137.ch10},
       adsurl = {https://ui.adsabs.harvard.edu/abs/2003and..book..345B}
}

@ARTICLE{berger2018poloidal,
       author = {{Berger}, M.~A. and {Hornig}, G.},
        title = "{A generalized poloidal-toroidal decomposition and an absolute measure of helicity}",
      journal = {Journal of Physics A Mathematical General},
         year = 2018,
        month = dec,
       volume = {51},
       number = {49},
          eid = {495501},
        pages = {495501},
          doi = {10.1088/1751-8121/aaea88},
       adsurl = {https://ui.adsabs.harvard.edu/abs/2018JPhA...51W5501B}
}

@ARTICLE{blackman2015ssr,
       author = {{Blackman}, Eric G.},
        title = "{Magnetic Helicity and Large Scale Magnetic Fields: A Primer}",
      journal = {\ssr},
         year = 2015,
        month = may,
       volume = {188},
       number = {1-4},
        pages = {59-91},
          doi = {10.1007/s11214-014-0038-6},
archivePrefix = {arXiv},
       eprint = {1402.0933},
 primaryClass = {astro-ph.SR},
       adsurl = {https://ui.adsabs.harvard.edu/abs/2015SSRv..188...59B}
}

@ARTICLE{bobra2014sharps,
       author = {{Bobra}, M.~G. and {Sun}, X. and {Hoeksema}, J.~T. and {Turmon}, M. and {Liu}, Y. and {Hayashi}, K. and {Barnes}, G. and {Leka}, K.~D.},
        title = "{The Helioseismic and Magnetic Imager (HMI) Vector Magnetic Field Pipeline: SHARPs - Space-Weather HMI Active Region Patches}",
      journal = {\solphys},
         year = 2014,
        month = sep,
       volume = {289},
       number = {9},
        pages = {3549-3578},
          doi = {10.1007/s11207-014-0529-3},
archivePrefix = {arXiv},
       eprint = {1404.1879},
 primaryClass = {astro-ph.SR},
       adsurl = {https://ui.adsabs.harvard.edu/abs/2014SoPh..289.3549B}
}

@ARTICLE{cantarella2000curl-spectrum,
       author = {{Cantarella}, Jason and {DeTurck}, Dennis and {Gluck}, Herman and {Teytel}, Mikhail},
        title = "{The spectrum of the curl operator on spherically symmetric domains}",
      journal = {Physics of Plasmas},
         year = 2000,
        month = jul,
       volume = {7},
       number = {7},
        pages = {2766-2775},
          doi = {10.1063/1.874127},
       adsurl = {https://ui.adsabs.harvard.edu/abs/2000PhPl....7.2766C}
}

@ARTICLE{candelaresi2011knots,
       author = {{Candelaresi}, Simon and {Brandenburg}, Axel},
        title = "{Decay of helical and nonhelical magnetic knots}",
      journal = {\pre},
         year = 2011,
        month = jul,
       volume = {84},
       number = {1},
          eid = {016406},
        pages = {016406},
          doi = {10.1103/PhysRevE.84.016406},
archivePrefix = {arXiv},
       eprint = {1103.3518},
 primaryClass = {astro-ph.SR},
       adsurl = {https://ui.adsabs.harvard.edu/abs/2011PhRvE..84a6406C}
}

@ARTICLE{candelaresi2021self-mutualwinding,
       author = {{Candelaresi}, Simon and {Hornig}, Gunnar and {MacTaggart}, David and {Simitev}, Radostin},
        title = "{On self and mutual winding helicity}",
      journal = {Commun Nonlinear Sci Numer Simulat},
         year = 2021,
       volume = {103},
          eid = {106015},
        pages = {106015},
          doi = {10.1016/j.cnsns.2021.106015}
}

@INPROCEEDINGS{cantarella2001bounds,
    author = {{Cantarella}, J. and {DeTurck}, D. and {Gluck}, H.},
    title = "{Upper Bounds for the Writhing of Knots and the Helicity of Vector Fields}",
    pages = {1--22},
    booktitle={Proceedings of the Conference in Honor of the 70th Birthday of Joan Birman},
    editor={{Gilman}, J. and {Lin}, X.-S. and {Menasco}, W.},
    year={2001},
    publisher={Americal Mathematical Society/International Press},
    address={Providence, RI}
}

@ARTICLE{demoulin2006mutualhelicity,
       author = {{Demoulin}, P. and {Pariat}, E. and {Berger}, M.~A.},
        title = "{Basic Properties of Mutual Magnetic Helicity}",
      journal = {\solphys},
         year = 2006,
        month = jan,
       volume = {233},
       number = {1},
        pages = {3-27},
          doi = {10.1007/s11207-006-0010-z},
       adsurl = {https://ui.adsabs.harvard.edu/abs/2006SoPh..233....3D}
}

@ARTICLE{demoulin2007review,
       author = {{D{\'e}moulin}, P.},
        title = "{Recent theoretical and observational developments in magnetic helicity studies}",
      journal = {Advances in Space Research},
         year = 2007,
        month = jan,
       volume = {39},
       number = {11},
        pages = {1674-1693},
          doi = {10.1016/j.asr.2006.12.037},
       adsurl = {https://ui.adsabs.harvard.edu/abs/2007AdSpR..39.1674D}
}

@ARTICLE{demoulin2009review,
       author = {{D{\'e}moulin}, P. and {Pariat}, E.},
        title = "{Modelling and observations of photospheric magnetic helicity}",
      journal = {Advances in Space Research},
         year = 2009,
        month = apr,
       volume = {43},
       number = {7},
        pages = {1013-1031},
          doi = {10.1016/j.asr.2008.12.004},
       adsurl = {https://ui.adsabs.harvard.edu/abs/2009AdSpR..43.1013D}
}

@ARTICLE{devore2000diffrot,
       author = {{DeVore}, C. Richard},
        title = "{Magnetic Helicity Generation by Solar Differential Rotation}",
      journal = {\apj},
         year = 2000,
        month = aug,
       volume = {539},
       number = {2},
        pages = {944-953},
          doi = {10.1086/309274},
       adsurl = {https://ui.adsabs.harvard.edu/abs/2000ApJ...539..944D}
}

@ARTICLE{durrant1989lfff,
       author = {{Durrant}, C.~J.},
        title = "{Linear force-free magnetic fields and coronal models}",
      journal = {Australian Journal of Physics},
         year = 1989,
        month = jan,
       volume = {42},
       number = {3},
        pages = {317-329},
          doi = {10.1071/PH890317},
       adsurl = {https://ui.adsabs.harvard.edu/abs/1989AuJPh..42..317D}
}

@ARTICLE{finn1985helicity,
       author = {{Finn}, John M. and {Antonsen}, Jr., Thomas M.},
        title = "{Magnetic helicity: What is it and what is it good for?}",
      journal = {Comments on Plasma Physics and Controlled Fusion},
         year = 1985,
        month = may,
       volume = {9},
        pages = {111-126},
       adsurl = {https://ui.adsabs.harvard.edu/abs/1985CoPPC...9..111F}
}

@ARTICLE{freedman1991,
       author = {{Freedman}, M.~H. and {He}, Z.-X.},
        title = "{Divergence-free fields: Energy and asymptotic crossing number}",
      journal = {Annals of Mathematics},
         year = 1991,
       volume = {134},
        pages = {189-229}
}

@ARTICLE{linan2018,
       author = {{Linan}, L. and {Pariat}, {\'E}. and {Moraitis}, K. and {Valori}, G. and {Leake}, J.},
        title = "{Time Variations of the Nonpotential and Volume-threading Magnetic Helicities}",
      journal = {\apj},
         year = 2018,
        month = sep,
       volume = {865},
       number = {1},
          eid = {52},
        pages = {52},
          doi = {10.3847/1538-4357/aadae7},
archivePrefix = {arXiv},
       eprint = {1809.03765},
 primaryClass = {astro-ph.SR},
       adsurl = {https://ui.adsabs.harvard.edu/abs/2018ApJ...865...52L}
}

@ARTICLE{mactaggart2023relativemultiply-connected,
       author = {{MacTaggart}, David and {Valli}, Alberto},
        title = "{Relative magnetic helicity in multiply connected domains}",
      journal = {Journal of Physics A Mathematical General},
         year = 2023,
        month = oct,
       volume = {56},
       number = {43},
          eid = {435701},
        pages = {435701},
          doi = {10.1088/1751-8121/acfd6c},
archivePrefix = {arXiv},
       eprint = {2307.14159},
 primaryClass = {physics.plasm-ph},
       adsurl = {https://ui.adsabs.harvard.edu/abs/2023JPhA...56Q5701M}
}

@ARTICLE{mactaggart2025bounds,
       author = {{MacTaggart}, David},
        title = "{A bound for relative magnetic helicity in terms of free magnetic energy}",
      journal = {Discover Space},
         year = 2025,
        month = dec,
       volume = {129},
       number = {1},
          eid = {19},
        pages = {19},
          doi = {10.1007/s11038-025-09578-8},
       adsurl = {https://ui.adsabs.harvard.edu/abs/2025DiSpa.129...19M}
}

@ARTICLE{martin1998,
       author = {{Martin}, Sara F.},
        title = "{Conditions for the Formation and Maintenance of Filaments   (Invited Review)}",
      journal = {\solphys},
         year = 1998,
        month = sep,
       volume = {182},
       number = {1},
        pages = {107-137},
          doi = {10.1023/A:1005026814076},
       adsurl = {https://ui.adsabs.harvard.edu/abs/1998SoPh..182..107M}
}

@ARTICLE{matthaeus1982solarwind,
       author = {{Matthaeus}, W.~H. and {Goldstein}, M.~L.},
        title = "{Measurement of the rugged invariants of magnetohydrodynamic turbulence in the solar wind}",
      journal = {\jgr},
         year = 1982,
        month = aug,
       volume = {87},
       number = {A8},
        pages = {6011-6028},
          doi = {10.1029/JA087iA08p06011},
       adsurl = {https://ui.adsabs.harvard.edu/abs/1982JGR....87.6011M}
}

@ARTICLE{moffatt1992,
       author = {{Moffatt}, H.~K. and {Ricca}, Renzo L.},
        title = "{Helicity and the Calugareanu Invariant}",
      journal = {Proceedings of the Royal Society of London Series A},
         year = 1992,
        month = nov,
       volume = {439},
       number = {1906},
        pages = {411-429},
          doi = {10.1098/rspa.1992.0159},
       adsurl = {https://ui.adsabs.harvard.edu/abs/1992RSPSA.439..411M}
}

@BOOK{moffatt2019book,
       author = {{Moffatt}, H.~K. and {Dormy}, E.},
        title = "{Self-Exciting Fluid Dynamos}",
         year = 2019,
          doi = {10.1017/9781107588691},
       adsurl = {https://ui.adsabs.harvard.edu/abs/2019sefd.book.....M}
}

@ARTICLE{moraitis2018comp,
       author = {{Moraitis}, Kostas and {Pariat}, {\'E}tienne and {Savcheva}, Antonia and {Valori}, Gherardo},
        title = "{Computation of Relative Magnetic Helicity in Spherical Coordinates}",
      journal = {\solphys},
         year = 2018,
        month = jun,
       volume = {293},
       number = {6},
          eid = {92},
        pages = {92},
          doi = {10.1007/s11207-018-1314-5},
archivePrefix = {arXiv},
       eprint = {1806.03011},
 primaryClass = {astro-ph.SR},
       adsurl = {https://ui.adsabs.harvard.edu/abs/2018SoPh..293...92M}
}

@INPROCEEDINGS{narita2024review,
       author = {{Narita}, Yasuhito},
        title = "{Magnetic Helicity Measurements in the Solar Wind}",
    booktitle = {Helicities in Geophysics, Astrophysics, and Beyond},
         year = 2024,
       editor = {{Kuzanyan}, Kirill and {Yokoi}, Nobumitsu and {Georgoulis}, Manolis K. and {Stepanov}, Rodion},
       volume = {283},
        month = jan,
        pages = {105-116},
          doi = {10.1002/9781119841715.ch7},
       adsurl = {https://ui.adsabs.harvard.edu/abs/2024GMS...283..105N}
}

@ARTICLE{pevtsov2003,
       author = {{Pevtsov}, A.~A. and {Balasubramaniam}, K.~S.},
        title = "{Helicity patterns on the sun}",
      journal = {Advances in Space Research},
         year = 2003,
        month = jan,
       volume = {32},
       number = {10},
        pages = {1867-1874},
          doi = {10.1016/S0273-1177(03)90620-X},
       adsurl = {https://ui.adsabs.harvard.edu/abs/2003AdSpR..32.1867P}
}

@ARTICLE{pevtsov2014review,
       author = {{Pevtsov}, Alexei A. and {Berger}, Mitchell A. and {Nindos}, Alexander and {Norton}, Aimee A. and {van Driel-Gesztelyi}, Lidia},
        title = "{Magnetic Helicity, Tilt, and Twist}",
      journal = {\ssr},
         year = 2014,
        month = dec,
       volume = {186},
       number = {1-4},
        pages = {285-324},
          doi = {10.1007/s11214-014-0082-2},
       adsurl = {https://ui.adsabs.harvard.edu/abs/2014SSRv..186..285P}
}

@ARTICLE{pontin2016braids,
       author = {{Pontin}, D.~I. and {Candelaresi}, S. and {Russell}, A.~J.~B. and {Hornig}, G.},
        title = "{Braided magnetic fields: equilibria, relaxation and heating}",
      journal = {Plasma Physics and Controlled Fusion},
         year = 2016,
        month = may,
       volume = {58},
       number = {5},
          eid = {054008},
        pages = {054008},
          doi = {10.1088/0741-3335/58/5/054008},
archivePrefix = {arXiv},
       eprint = {1512.05918},
 primaryClass = {physics.plasm-ph},
       adsurl = {https://ui.adsabs.harvard.edu/abs/2016PPCF...58e4008P}
}

@ARTICLE{prior2014winding-gauge,
       author = {{Prior}, C. and {Yeates}, A.~R.},
        title = "{On the Helicity of Open Magnetic Fields}",
      journal = {\apj},
         year = 2014,
        month = jun,
       volume = {787},
       number = {2},
          eid = {100},
        pages = {100},
          doi = {10.1088/0004-637X/787/2/100},
archivePrefix = {arXiv},
       eprint = {1404.3897},
 primaryClass = {astro-ph.SR},
       adsurl = {https://ui.adsabs.harvard.edu/abs/2014ApJ...787..100P}
}

@ARTICLE{ricca2013,
       author = {{Ricca}, Renzo L.},
        title = "{New energy and helicity bounds for knotted and braided magnetic fields}",
      journal = {Geophysical and Astrophysical Fluid Dynamics},
         year = 2013,
        month = aug,
       volume = {107},
       number = {4},
        pages = {385-402},
          doi = {10.1080/03091929.2012.681782},
       adsurl = {https://ui.adsabs.harvard.edu/abs/2013GApFD.107..385R}
}

@ARTICLE{schuck2024linkages,
       author = {{Schuck}, Peter W. and {Linton}, Mark G.},
        title = "{Disentangling the Entangled Linkages of Relative Magnetic Helicity}",
      journal = {\apj},
         year = 2024,
        month = feb,
       volume = {961},
       number = {2},
          eid = {156},
        pages = {156},
          doi = {10.3847/1538-4357/acf471},
archivePrefix = {arXiv},
       eprint = {2309.07776},
 primaryClass = {astro-ph.SR},
       adsurl = {https://ui.adsabs.harvard.edu/abs/2024ApJ...961..156S}
}

@ARTICLE{thalmann2021issi4,
       author = {{Thalmann}, J.~K. and {Georgoulis}, M.~K. and {Liu}, Y. and {Pariat}, E. and {Valori}, G. and {Anfinogentov}, S. and {Chen}, F. and {Guo}, Y. and {Moraitis}, K. and {Yang}, S. and {Mastrano}, Alpha and {ISSI Team on Magnetic Helicity}},
        title = "{Magnetic Helicity Estimations in Models and Observations of the Solar Magnetic Field. IV. Application to Solar Observations}",
      journal = {\apj},
         year = 2021,
        month = nov,
       volume = {922},
       number = {1},
          eid = {41},
        pages = {41},
          doi = {10.3847/1538-4357/ac1f93},
archivePrefix = {arXiv},
       eprint = {2108.08525},
 primaryClass = {astro-ph.SR},
       adsurl = {https://ui.adsabs.harvard.edu/abs/2021ApJ...922...41T}
}

@ARTICLE{toriumi2019lrsp,
       author = {{Toriumi}, Shin and {Wang}, Haimin},
        title = "{Flare-productive active regions}",
      journal = {Living Reviews in Solar Physics},
         year = 2019,
        month = dec,
       volume = {16},
       number = {1},
          eid = {3},
        pages = {3},
          doi = {10.1007/s41116-019-0019-7},
archivePrefix = {arXiv},
       eprint = {1904.12027},
 primaryClass = {astro-ph.SR},
       adsurl = {https://ui.adsabs.harvard.edu/abs/2019LRSP...16....3T}
}

@ARTICLE{valori2016issi,
       author = {{Valori}, Gherardo and {Pariat}, Etienne and {Anfinogentov}, Sergey and {Chen}, Feng and {Georgoulis}, Manolis K. and {Guo}, Yang and {Liu}, Yang and {Moraitis}, Kostas and {Thalmann}, Julia K. and {Yang}, Shangbin},
        title = "{Magnetic Helicity Estimations in Models and Observations of the Solar Magnetic Field. Part I: Finite Volume Methods}",
      journal = {\ssr},
         year = 2016,
        month = nov,
       volume = {201},
       number = {1-4},
        pages = {147-200},
          doi = {10.1007/s11214-016-0299-3},
archivePrefix = {arXiv},
       eprint = {1610.02193},
 primaryClass = {astro-ph.SR},
       adsurl = {https://ui.adsabs.harvard.edu/abs/2016SSRv..201..147V}
}

@ARTICLE{valori2020additivity,
       author = {{Valori}, Gherardo and {D{\'e}moulin}, Pascal and {Pariat}, Etienne and {Yeates}, Anthony and {Moraitis}, Kostas and {Linan}, Luis},
        title = "{Additivity of relative magnetic helicity in finite volumes}",
      journal = {\aap},
         year = 2020,
        month = nov,
       volume = {643},
          eid = {A26},
        pages = {A26},
          doi = {10.1051/0004-6361/202038533},
archivePrefix = {arXiv},
       eprint = {2008.00968},
 primaryClass = {astro-ph.SR},
       adsurl = {https://ui.adsabs.harvard.edu/abs/2020A&A...643A..26V}
}

@ARTICLE{woltjer1958theorem,
       author = {{Woltjer}, L.},
        title = "{A Theorem on Force-Free Magnetic Fields}",
      journal = {Proceedings of the National Academy of Science},
         year = 1958,
        month = jun,
       volume = {44},
       number = {6},
        pages = {489-491},
          doi = {10.1073/pnas.44.6.489},
       adsurl = {https://ui.adsabs.harvard.edu/abs/1958PNAS...44..489W}
}

@ARTICLE{xiao2023spherical,
       author = {{Xiao}, Daining and {Prior}, Christopher B. and {Yeates}, Anthony R.},
        title = "{Spherical winding and helicity}",
      journal = {Journal of Physics A Mathematical General},
         year = 2023,
        month = may,
       volume = {56},
       number = {20},
          eid = {205201},
        pages = {205201},
          doi = {10.1088/1751-8121/accc17},
       adsurl = {https://ui.adsabs.harvard.edu/abs/2023JPhA...56t5201X}
}

@ARTICLE{xiao2025periodic,
       author = {{Xiao}, Daining and {Prior}, Christopher B. and {Yeates}, Anthony R.},
        title = "{Winding and magnetic helicity in periodic domains}",
      journal = {Proceedings of the Royal Society of London Series A},
         year = 2025,
        month = feb,
       volume = {481},
       number = {2307},
          eid = {20240152},
        pages = {20240152},
          doi = {10.1098/rspa.2024.0152},
       adsurl = {https://ui.adsabs.harvard.edu/abs/2025RSPSA.48140152X}
}

@ARTICLE{yeates2014bianchi,
       author = {{Yeates}, A.~R. and {Bianchi}, F. and {Welsch}, B.~T. and {Bushby}, P.~J.},
        title = "{The coronal energy input from magnetic braiding}",
      journal = {\aap},
         year = 2014,
        month = apr,
       volume = {564},
          eid = {A131},
        pages = {A131},
          doi = {10.1051/0004-6361/201323276},
archivePrefix = {arXiv},
       eprint = {1403.4396},
 primaryClass = {astro-ph.SR},
       adsurl = {https://ui.adsabs.harvard.edu/abs/2014A&A...564A.131Y}
}

@ARTICLE{yeates2016global-helicity,
       author = {{Yeates}, A.~R. and {Hornig}, G.},
        title = "{The global distribution of magnetic helicity in the solar corona}",
      journal = {\aap},
         year = 2016,
        month = oct,
       volume = {594},
          eid = {A98},
        pages = {A98},
          doi = {10.1051/0004-6361/201629122},
archivePrefix = {arXiv},
       eprint = {1606.06863},
 primaryClass = {astro-ph.SR},
       adsurl = {https://ui.adsabs.harvard.edu/abs/2016A&A...594A..98Y}
}

@ARTICLE{yeates2018issi,
       author = {{Yeates}, Anthony R. and {Amari}, Tahar and {Contopoulos}, Ioannis and {Feng}, Xueshang and {Mackay}, Duncan H. and {Miki{\'c}}, Zoran and {Wiegelmann}, Thomas and {Hutton}, Joseph and {Lowder}, Christopher A. and {Morgan}, Huw and {Petrie}, Gordon and {Rachmeler}, Laurel A. and {Upton}, Lisa A. and {Canou}, Aurelien and {Chopin}, Pierre and {Downs}, Cooper and {Druckm{\"u}ller}, Miloslav and {Linker}, Jon A. and {Seaton}, Daniel B. and {T{\"o}r{\"o}k}, Tibor},
        title = "{Global Non-Potential Magnetic Models of the Solar Corona During the March 2015 Eclipse}",
      journal = {\ssr},
         year = 2018,
        month = aug,
       volume = {214},
       number = {5},
          eid = {99},
        pages = {99},
          doi = {10.1007/s11214-018-0534-1},
archivePrefix = {arXiv},
       eprint = {1808.00785},
 primaryClass = {astro-ph.SR},
       adsurl = {https://ui.adsabs.harvard.edu/abs/2018SSRv..214...99Y}
}

@ARTICLE{yeates2020minimal-helicity,
       author = {{Yeates}, Anthony R.},
        title = "{The Minimal Helicity of Solar Coronal Magnetic Fields}",
      journal = {\apjl},
         year = 2020,
        month = aug,
       volume = {898},
       number = {2},
          eid = {L49},
        pages = {L49},
          doi = {10.3847/2041-8213/aba762},
archivePrefix = {arXiv},
       eprint = {2007.10649},
 primaryClass = {astro-ph.SR},
       adsurl = {https://ui.adsabs.harvard.edu/abs/2020ApJ...898L..49Y}
}

@ARTICLE{yeates2024cycle24,
       author = {{Yeates}, Anthony R.},
        title = "{The Sun's Non-Potential Corona over Solar Cycle 24}",
      journal = {\solphys},
         year = 2024,
        month = jun,
       volume = {299},
       number = {6},
          eid = {83},
        pages = {83},
          doi = {10.1007/s11207-024-02328-5},
archivePrefix = {arXiv},
       eprint = {2405.14322},
 primaryClass = {astro-ph.SR},
       adsurl = {https://ui.adsabs.harvard.edu/abs/2024SoPh..299...83Y}
}

@INPROCEEDINGS{yeates2024flhreview,
       author = {{Yeates}, Anthony R. and {Berger}, Mitchell A.},
        title = "{Introduction to Field Line Helicity}",
    booktitle = {Helicities in Geophysics, Astrophysics, and Beyond},
         year = 2024,
       editor = {{Kuzanyan}, Kirill and {Yokoi}, Nobumitsu and {Georgoulis}, Manolis K. and {Stepanov}, Rodion},
       volume = {283},
        month = jan,
        pages = {1-16},
          doi = {10.1002/9781119841715.ch1},
archivePrefix = {arXiv},
       eprint = {2311.08831},
 primaryClass = {astro-ph.SR},
       adsurl = {https://ui.adsabs.harvard.edu/abs/2024GMS...283....1Y}
}

@ARTICLE{yi2022poloidal,
       author = {{Yi}, Sibaek and {Choe}, G.~S.},
        title = "{The toroidal field surfaces in the standard poloidal-toroidal representation of magnetic field}",
      journal = {Scientific Reports},
         year = 2022,
        month = feb,
       volume = {12},
          eid = {2944},
        pages = {2944},
          doi = {10.1038/s41598-022-07040-7},
archivePrefix = {arXiv},
       eprint = {2112.13665},
 primaryClass = {physics.class-ph},
       adsurl = {https://ui.adsabs.harvard.edu/abs/2022NatSR..12.2944Y}
}

@ARTICLE{yoshida1990rot,
       author = {{Yoshida}, Zensho and {Giga}, Yoshikazu},
        title = "{Remarks of Spectra of Operator Rot}",
      journal = {Math. Z.},
         year = 1990,
       volume = {204},
        pages = {235-245}
}

@misc{sphtools-code,
  author    = {{Yeates}, Anthony R.},
  title     = {sphtools},
  version   = {v1.0},
  year      = {2025},
  publisher = {Zenodo},
  doi       = {10.5281/zenodo.17607526},
  url       = {https://doi.org/10.5281/zenodo.17607526}
}
\bibliographystyle{aasjournal}

\end{document}